\documentclass{article}

\usepackage{microtype}
\usepackage{graphicx}
\usepackage{subcaption}
\usepackage{booktabs}
\usepackage[dvipsnames]{xcolor}
\usepackage{tikz}
\usepackage[notransparent]{svg}
\usepackage{multirow}
\usepackage{xurl}
\usepackage{enumitem}
\usepackage{balance}

\usepackage{xcolor}

\usepackage{hyperref}

\usetikzlibrary{
    positioning,
    backgrounds, 
    fit, 
    calc,}

\definecolor{blue}{HTML}{1f77b4}
\definecolor{red}{HTML}{d62728}
\definecolor{green}{HTML}{2ca02c}
\definecolor{orange}{HTML}{ff7f0e}
\definecolor{purple}{HTML}{9467bd}
\definecolor{brown}{HTML}{8c564b}
\definecolor{pink}{HTML}{e377c2}
\definecolor{gray}{HTML}{7f7f7f}
\definecolor{olive}{HTML}{bcbd22}
\definecolor{cyan}{HTML}{17becf}

\tikzset{bb/.style={draw, inner sep=1.mm, rounded corners}}

\newcommand*\circled[1]{\protect\tikz[baseline=(char.base)]{\protect\node[shape=circle,fill=black,inner sep=0.6pt] (char) {\textcolor{white}{\footnotesize {#1}}};}}

\usepackage[accepted]{icml2025}

\usepackage{amsmath}
\usepackage{amssymb}
\usepackage{mathtools}
\usepackage{amsthm}
\usepackage{array} 

\usepackage[capitalize,noabbrev]{cleveref}

\newcommand{\pmS}[1]{
    {\fontsize{7}{7}\selectfont $\pm$ #1}
    \vspace{0.007cm}
}

\renewcommand{\paragraph}[1]{\textbf{#1}}

\theoremstyle{plain}

\theoremstyle{definition}

\theoremstyle{remark}

\usepackage[textsize=tiny]{todonotes}

\icmltitlerunning{Benchmarking Self-Supervised Learning for Single-Cell Data}

\begin{document}

\twocolumn[
\icmltitle{scSSL-Bench: Benchmarking Self-Supervised Learning for Single-Cell Data}



\icmlsetsymbol{equal}{*}
\icmlsetsymbol{equal_last}{$\dagger$}

\begin{icmlauthorlist}
\icmlauthor{Olga Ovcharenko}{equal,TUB}
\icmlauthor{Florian Barkmann}{equal,ETH}
\icmlauthor{Philip Toma}{equal,ETH}
\icmlauthor{Imant Daunhawer}{ETH}
\icmlauthor{Julia E. Vogt}{ETH}
\icmlauthor{Sebastian Schelter}{equal_last,TUB}
\icmlauthor{Valentina Boeva}{equal_last,ETH,SIB,Inserm}
\end{icmlauthorlist}

\icmlaffiliation{ETH}{Department of Computer Science, ETH Zurich, Zurich, Switzerland}
\icmlaffiliation{TUB}{BIFOLD \& TU Berlin, Berlin, Germany}
\icmlaffiliation{SIB}{Swiss Institute of Bioinformatics, Lausanne, Switzerland}
\icmlaffiliation{Inserm}{Paris Cité University, Cochin Institute, INSERM U1016, Paris, France}

\icmlcorrespondingauthor{Olga Ovcharenko}{ovcharenko@tu-berlin.de}
\icmlcorrespondingauthor{Sebastian Schelter}{schelter@tu-berlin.de}
\icmlcorrespondingauthor{Valentina Boeva}{valentina.boeva@inf.ethz.ch}

\icmlkeywords{Machine Learning, ICML}

\vskip 0.3in
]



\printAffiliationsAndNotice{\textsuperscript{*$\dagger$} Equal contribution} 

\begin{abstract}
Self-supervised learning (SSL) has proven to be a powerful approach for extracting biologically meaningful representations from single-cell data. To advance our understanding of SSL methods applied to single-cell data, we present scSSL-Bench, a comprehensive benchmark that evaluates nineteen SSL methods. Our evaluation spans nine datasets and focuses on three common downstream tasks: batch correction, cell type annotation, and missing modality prediction.
Furthermore, we systematically assess various data augmentation strategies.
Our analysis reveals task-specific trade-offs: the specialized single-cell frameworks, scVI, CLAIRE, and the finetuned scGPT excel at uni-modal batch correction, while generic SSL methods, such as VICReg and SimCLR, demonstrate superior performance in cell typing and multi-modal data integration. Random masking emerges as the most effective augmentation technique across all tasks, surpassing domain-specific augmentations. Notably, our results indicate the need for a specialized single-cell multi-modal data integration framework. scSSL-Bench provides a standardized evaluation platform and concrete recommendations for applying SSL to single-cell analysis, advancing the convergence of deep learning and single-cell genomics.


\end{abstract}

\vspace{-0.5cm}

\section{Introduction}
\label{sec:introduction}

Recent progress in single-cell RNA sequencing (scRNA-seq) and multi-omics sequencing technologies has transformed our understanding of cellular heterogeneity by enabling cell molecular profiling at unprecedented resolution~\citep{sikkema2022integrated,eraslan2022single}. This breakthrough has revolutionized our ability to understand diseases, develop personalized treatments, and trace the origins of complex conditions like cancer and autoimmune disorders. scRNA-seq~\cite{scRNA-seq} captures gene expression levels in individual cells and generates a high-dimensional matrix where each row represents a cell and each column represents a gene's expression level. Multi-omics approaches simultaneously measure additional molecular features, including chromatin accessibility through ATAC-seq~\cite{ATAC-seq} or protein levels via CITE-seq~\cite{CITE-seq}. Modern multi-omics experiments generate massive datasets encompassing hundreds of thousands of cells, with each cell characterized by diverse measurements: the expression of tens of thousands of genes, the accessibility of hundreds of thousands of chromatin regions, and the abundance of hundreds of surface proteins. Multi-modal profiling provides an unprecedented view of cellular state and function. However, the resulting datasets are susceptible to \emph{batch effects} — systematic technical variations introduced during sample preparation, sequencing, or processing~\cite{lahnemann2020eleven}. If left uncorrected, batch effects mask genuine biological signals and compromise downstream analyses~\cite{heumos2023best}. For instance, when comparing blood samples from cancer patients processed in different laboratories, batch effects can make immune cells from the same patient appear more different from each other than from cells of other patients, masking crucial patterns in how the immune system responds to the tumor~\citep{slyper2020single}. 

The success of self-supervised learning (SSL) methods in computer vision~\citep{he2020momentum, chen2020simple}, video processing~\cite{schiappa2023self}, and natural language processing~\cite{min2023recent} has inspired their application to single-cell data. Several models have been adapted for analyzing single-cell data~\cite{CLEAR, liu2024cake, scBridge, monae}, showing promising results in mitigating batch effects and improving downstream analyses. 
There is an interest in the genomics community in finding standardized approaches for applying SSL to single-cell analysis.
A recent work~\cite{Richter2024} discusses scenarios in which SSL is applicable to single-cell genomics. The authors compare the performance of masked autoencoders and two SSL methods (BYOL~\cite{grill2020bootstrap} and Barlow~Twins~\cite{zbontar2021barlowtwinsselfsupervisedlearning}), discuss the effects of pre-training on auxiliary data, and empirically study the efficacy of zero-shot and fine-tuned SSL. However, their work lacks a comparison to specialized single-cell models and does not explore individual hyperparameters and regularization techniques.
Furthermore, building upon innovations in natural language processing~\cite{vaswani2017attention, devlin2019bert}, single-cell foundation models~\cite{cui2024scgpt, yang2022scbert, theodoris2023transfer} have recently emerged as powerful tools to understand cellular heterogeneity and gene-gene interactions, and require a comparison to contrastive methods.

Since the majority of SSL methods were originally developed for image and text data, there is a lack of a systematic comparison of models, hyperparameters, training regimes, regularization techniques, and augmentations for single-cell genomics data~\cite{toma2024benchmarking}.
This knowledge gap limits our understanding of how to effectively adapt and optimize SSL methods for single-cell data. Our work seeks to fill this gap and focuses on the following research questions:

\vspace{-0.3cm}
\begin{itemize}[leftmargin=*]
    \item \emph{RQ1} -- Do specialized single-cell SSL methods outperform generic SSL methods? How does the performance of SSL models differ for uni-omics and multi-omics data? 
    \item \emph{RQ2} -- How do hyperparameters and augmentation techniques impact the performance of generic SSL methods for single-cell data?
    \item \emph{RQ3} -- Are batch normalization and multi-modal integration techniques proposed for image data beneficial for single-cell genomics data as well?
\end{itemize}
\vspace{-0.3cm}

Our main contribution is an open-source benchmark, scSSL-Bench, which compares the performance of several self-supervised learning methods for single-cell data.
(1) To address \emph{RQ1}, we evaluate nineteen generic and specialized single-cell SSL methods across seven different single-cell uni-modal and two multi-modal datasets, assessing their performance on three common downstream tasks: batch correction, cell type annotation, and missing modality prediction (\autoref{sec:experiments_tasks}). 
Our results reveal that specialized frameworks, scVI and CLAIRE, together with the foundation model, scGPT, are the best for uni-modal batch correction, while generic SSL techniques such as VICReg and SimCLR outperform domain-specific methods for multi-modal batch correction and the other two tasks on single-modal data.
(2) For \emph{RQ2}, we evaluate various model architectures and hyperparameters, including representation and projection dimensionality, augmentation strategies, and multi-modal integration methods. (\autoref{sec:experiments_hyperparam}).
Overall, we find that a moderate to larger embedding dimensionality consistently leads to improved results and identify masking as the most beneficial augmentation technique that surpasses biology-specific augmentations.
(3) An assessment of design decisions suggested in the related work, e.g., retaining projector and domain-specific batch normalization, helps to answer \emph{RQ3} and to find best practices that can be adopted by the single-cell genomics community (\autoref{sec:experiments:common_com_practices}).
We find that neither domain-specific batch normalization nor retaining the projector during inference improves results.

We provide our benchmark code under an open license at {\small \url{https://github.com/BoevaLab/scSSL-Bench}}  for reproducibility and for fostering further research on SSL for single-cell data.

\section{Background}
\label{sec:background}

In this Section, we discuss SSL on single-cell data, the corresponding downstream tasks, and specialized methods.

\subsection{Machine Learning on Single-Cell Data}
\label{sec:background_tasks}

\paragraph{Data:} 
There exist multiple technologies that measure different aspects of the cellular state.
scRNA-seq~\cite{scRNA-seq} measures which genes are expressed in each cell and produces a high-dimensional sparse count matrix representing individual cells as rows and genes as columns.
CITE-seq~\cite{CITE-seq}, 10x multiome~\cite{Baysoy2023}, and TEA-seq~\cite{TEA-seq} profile complementary to gene expression aspects such as chromatin accessibility and protein abundance within a cell.

\paragraph{Downstream Tasks:}
Learned cell representations are commonly used for multiple downstream tasks.

\noindent {\em Batch correction} -- Single-cell data can be affected by batch effects, which challenges the ability to measure true biological variation~\cite{Yu2023, polanski2019bbknn}.
Batch effects are technical biases introduced while sequencing because of differences in sequencing platforms, timing, reagents, or experimental conditions across laboratories~\cite{Zhang2023DT}. 
To address the presence of batch effects, a common approach is learning a batch-corrected lower-dimensional embedding, where cells cluster based on their cell type and cell state rather than their experimental batch of origin~\cite{seurat} (\autoref{fig:TaskBatchCorrection} in the Appendix illustrates cells before and after batch correction).

\begin{figure*}[!t]
  \centering
  \includegraphics[width=\linewidth]{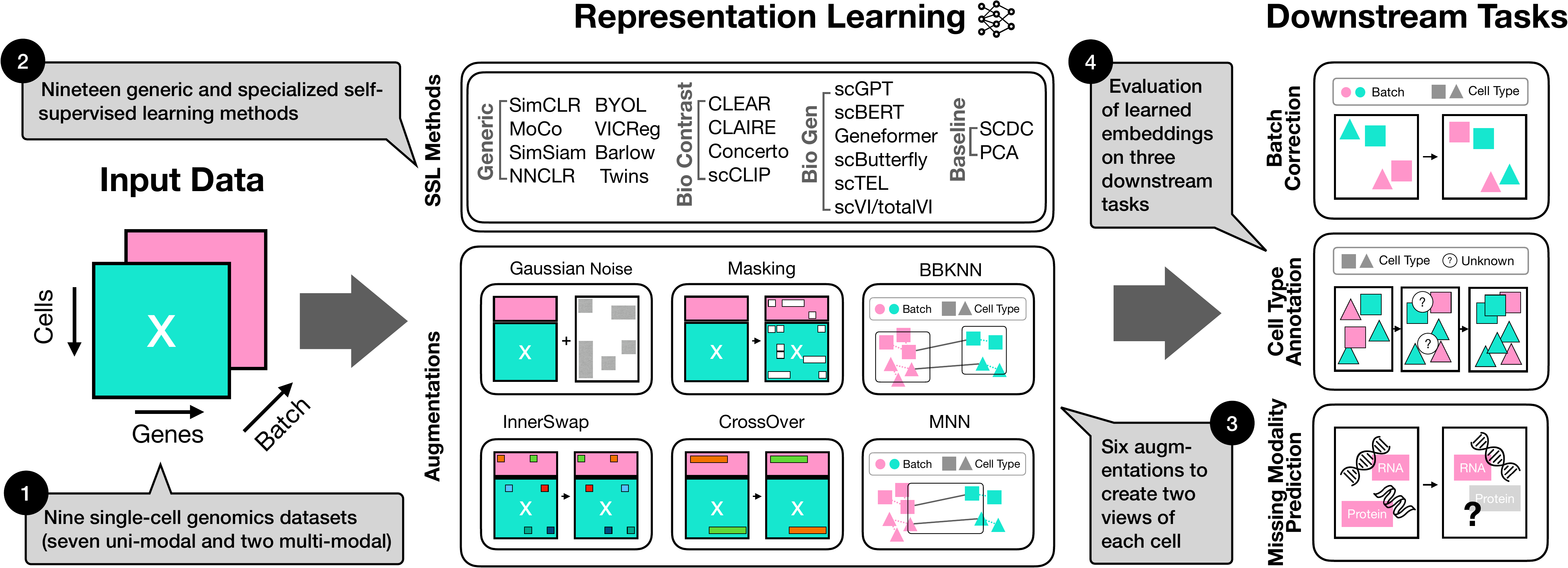}
    
  \vspace{-0.34cm}
  \caption{Outline of scSSL-Bench: \circled{1} As input, scSSL-Bench takes scRNA-seq data (cell-by-gene count matrix), where each value in the matrix represents the number of reads in a cell for the corresponding gene. \circled{2} scSSL-Bench trains one of nineteen methods: Generic, specialized contrastive (Bio Contrast), specialized generative (Bio Gen), and baselines. For self-supervised generic methods, scSSL-Bench uses augmentations \circled{3} to create two views of a cell. \circled{4} The learned embeddings are evaluated on three downstream tasks.}
  \label{fig:bench_design}
  
  \vspace{-0.2cm}
  
\end{figure*}

\noindent {\em Cell type annotation} -- This task (also called as query-to-reference mapping) revolves around unsupervised transfer learning~\cite{Concerto}, where the primary objective is to annotate cells of a hold-out dataset (query) by mapping them to a joint latent space of a pre-annotated train dataset (reference), whose cell types are known~\cite{Lotfollahi2022}. 
Once test and train data are aligned, 
held-out cells are annotated using a classifier trained on embeddings of the reference dataset.
\autoref{fig:TaskQueryToReference} in the Appendix visualizes how the learned representations of train and hold-out sets and train cell types are used to predict the cell types of hold-out data (query, blue result) during subsequent inference.

\noindent {\em Missing modality prediction} -- For multi-modal datasets, missing modality prediction enables the inference of unmeasured (missing) modalities in held-out (query) cells~\cite{Concerto}.
Given multi-modal train data (reference) with RNA and protein expressions and hold-out data containing only RNA, the goal is to predicts the hold-out dataset's original protein values by averaging the proteins of the nearest neighbors from the train set (referred to as kNN probing).

\subsection{Self-Supervised Learning (SSL) Methods}
\label{sec:backgroun_methods}

SSL aims to discover useful data representations without relying on annotations~\cite{geiping2023cookbook} by leveraging the dis-/similarity of data samples.
We refer to ~\autoref{sec:generic_methods} for details on generic SSL methods (and their contrastive and non-contrastive variants).

\paragraph{Single-Cell Contrastive Methods:} There are several SSL frameworks tailored for single-cell data.
CLEAR~\cite{CLEAR} employs contrastive SSL and leverages InfoNCE loss~\cite{infonce}. 
Positive/negative pairs are created by adding Gaussian noise, random masking, or  crossing over genes between two cells. 
CLAIRE~\cite{XuhuaRFL23} suggests a novel augmentation strategy by finding mutual nearest neighbors (MNN) between and nearest neighbors (KNN) within experimental batches in a dataset.
CLAIRE uses inter-biological-batch MNN pairs as initial positive pair seeds, which are then ``mixed'' with intra-biological-batch neighbors to generate positive pairs.
CLAIRE extends MoCo's~\cite{he2020momentum} architecture with online and momentum encoders.
Concerto~\cite{Concerto} is a contrastive self-supervised distillation framework that uses an asymmetric teacher-student network structure~\cite{hu2023teacherstudentarchitectureknowledgedistillation} and dropout to create two augmented views of a cell (a positive pair).
Positive/negative pairs are contrasted using NTXent loss~\cite{ntxent}. 
Concerto also supports single-cell multi-modal data, e.g., pairs of RNA and protein.
The scCLIP~\cite{scCLIP} method is a generalized multi-modal transformer model that applies contrastive learning to single-cell multi-omics data, which adopts ideas from CLIP~\cite{CLIP} by defining modality-specific encoders, constructing positive pairs from two modalities of the same cell, and contrasting them using InfoNCE loss~\cite{infonce}.

\paragraph{Single-Cell Generative Methods:} 
Generative approaches, from variational autoencoders (VAE) to transformer-based foundation models, are used to learn single-cell representations and exploit the biological batch/cell type annotations during training, which can be seen as leaking information compared to contrastive SSL methods.
The state-of-the-art single-cell method scVI~\cite{scvi} is a widely-used VAE that leverages a zero-inflated negative binomial distribution as reconstruction loss. 
A multi-modal version of scVI, totalVI~\cite{Gayoso2021}, allows joint analysis of RNA and protein expressions.
SCDC~\cite{SCDC} is a uni-modal method that employs biological batch and cell type encoders to create a concatenated representation that is reconstructed using the decoder. 
To improve the discrimination of the batch information, SCDC uses a specialized batch discriminator.
scTEL~\cite{scTEL} leverages transformer and LSTM layers to establish a mapping from RNA expression to unobserved protein expression in the same cells. 
scButterfly~\cite{scButterfly} supports CITE-seq~\cite{CITE-seq} data and employs a dual-VAE architecture with modality-specific pretraining and a discriminator to encourage the mixing of different modalities. 
Recently, transformer-based single-cell foundation models (scFMs) have emerged. 
For generalizability, scFMs are pre-trained on tens of millions of cells.
scBERT~\cite{yang2022scbert} adapts BERT's~\cite{devlin-etal-2019-bert} masked language modeling to learn contextual gene representations, scGPT~\cite{cui2024scgpt} uses GPT-style pretraining to create transferable representations across cell types and experimental conditions, and Geneformer~\cite{theodoris2023transfer} employs a transformer architecture pre-trained on large-scale gene expression datasets to capture gene-gene interactions.

\section{Benchmark Design}
\label{sec:benchmark-design}

\autoref{fig:bench_design} illustrates our design of scSSL-Bench. The input of the benchmark are cell-by-gene count matrices containing scRNA-seq or CITE-seq data~\circled{1}. Depending on the data and SSL method, scSSL-Bench trains one of nineteen representation learning frameworks~\circled{2} using augmentations to create positive/negative pairs~\circled{3} for self-supervised approaches. Finally, the learned representations are evaluated on three downstream tasks~\circled{4}.

\subsection{Datasets, Models, and Tasks}

\paragraph{Datasets:} We consider nine single-cell genomics datasets that represent common established benchmarks~\cite{Richter2024}. 
Peripheral Blood Mononuclear Cells (PBMC), Pancreas, Immune Cell Atlas, Mouse Cell Atlas (MCA), Human Immune Cells (HIC), Lung, and Tabula Sapiens are seven single-modal datasets collected using scRNA-seq~\cite{scRNA-seq} technology.
Multi-modal Peripheral Blood Mononuclear Cells (PBMC-M) and Multi-modal Bone Marrow Mononuclear Cells (BMMC) are multi-modal datasets collected using CITE-seq technology that contain RNA and protein or gene expression and protein abundance (ADT) respectively.
Further details in~\autoref{datasets}.

\paragraph{SSL Methods:} 
To investigate \emph{RQ1}, we benchmark nineteen existing SSL methods and divide them into four categories: generic, domain-specific specialized contrastive and generative methods, and baselines.
SimCLR~\cite{chen2020simple}, MoCo~\cite{he2020momentum}, SimSiam~\cite{chen2020exploringsimplesiameserepresentation},  NNCLR~\cite{dwibedi2021littlehelpfriendsnearestneighbor}, BYOL~\cite{grill2020bootstrap}, VICReg~\cite{vicreg}, and BarlowTwins~\cite{zbontar2021barlowtwinsselfsupervisedlearning} are generic SSL architectures that we adopt to single-cell (multi-omics) data (see~\autoref{fig:Model-Overview} for architecture details). 
Contrastive domain-specific methods that are tailored for the single-cell data include Concerto~\cite{Concerto}, CLEAR~\cite{CLEAR}, CLAIRE~\cite{XuhuaRFL23}, and scCLIP~\cite{scCLIP}. 
Generative methods include commonly used for single-cell data integration, scVI~\cite{scvi} and totalVI~\cite{Gayoso2021}, which are single-cell specialized variational autoencoder-based methods, and single-cell foundation models scGPT~\cite{cui2024scgpt}, Geneformer~\cite{theodoris2023transfer}, and scBERT~\cite{yang2022scbert}. 
Additionally, for multi-omics integration, we evaluate scButterfly~\cite{scButterfly} and scTEL~\cite{scTEL}, which leverage variational autoencoders.
SCDC~\cite{SCDC} and principal component analysis (PCA)~\cite{pearson1901liii} are used as baselines in scSSL-Bench. We include PCA as a baseline to assess whether more complex SSL methods offer substantial improvements over a simple linear dimensionality reduction technique that does not correct for batch effects.
We refer to~\autoref{sec:backgroun_methods} for detailed descriptions of each method.

\noindent First, in all contrastive methods except Concerto, two views are created by augmenting a single sample.
Second, both views are encoded by a network with shared weights, producing data representations.
Concerto removes the necessity for transforming samples by placing a dropout layer behind the encoder backbone.
Finally, while training, all representations produced by the encoder are passed into a projector to improve robustness~\cite{xue2024investigatingbenefitsprojectionhead}. 
In all contrastive approaches but Concerto and scCLIP, the projector is discarded during inference, keeping only the encoder's output.

\paragraph{Downstream Tasks and Evaluation:}
To address \emph{RQ1}, our benchmark evaluates multiple single-cell datasets on three tasks: batch correction, cell type annotation, and modality prediction (see~\autoref{sec:background_tasks} for details).

\noindent {\em Batch correction} -- the quality of batch-corrected embeddings is measured by biological conservation and batch correction metrics. These metrics were introduced in single-cell integration benchmarking (scIB)~\cite{Buttner2019, luecken2022benchmarking, Tran2020}, a tool that is widely used in the single-cell community, see~\autoref{sec:eval_details} for details.
Analogous to \citeauthor{luecken2022benchmarking} \citeyear{luecken2022benchmarking}, we combine bio conservation (Bio), measuring the similarity between cell embeddings and ground-truth cell types or states, and batch correction (Batch), measuring how well the batch effect is removed, by aggregating these scores into a total score by $Total = 0.6 \times Bio  + 0.4 \times Batch$.
All tables showing batch correction results are min-max scaled inside each dataset.

\noindent {\em Cell type annotation} -- each dataset is divided into train (reference) and test (query) data that consists of up to three held-out (experimental) batches with unseen cells (details in~\autoref{sec:eval_details}). 
We train a k-nearest neighbors (KNN) classifier with train (reference) embeddings and cell types to annotate test (query) data representation. Next, k-nearest neighbor probing~\cite{marks2024closerlookbenchmarkingselfsupervised} is used to predict cell types, and performance is evaluated using the macro-average F1-score and classification accuracy~\cite{Heryanto2024}.

\noindent {\em Missing modality prediction} on multi-modal datasets -- we evaluate the quality of the inferred modality by measuring the Pearson correlation between the original and predicted values, see \autoref{sec:eval_details} for more details.

\subsection{Augmentation, Batch Normalization, and Multi-Modal Integration}

\paragraph{Augmentations:} We evaluate augmentations for single-cell data proposed in CLEAR~\cite{CLEAR} and CLAIRE~\cite{XuhuaRFL23} to investigate \emph{RQ2}. The purpose of augmentations in contrastive SSL is to transform the original sample into two distinct views that are contrasted during training~\cite{zhang2022rethinkingaugmentationmodulecontrastive}. 
Multiple augmentations can be applied to a data sample to improve the generalization and robustness of representations.
The authors of CLEAR~\cite{CLEAR} introduce four augmentations for scRNA-seq data, each of which we apply with 50\% probability: Masking, Gaussian noise, InnerSwap, and CrossOver. First, a random mask sets 20\% of a cell's genes to zero, followed by additive Gaussian noise (with mean 0 and standard deviation 0.2) to 80\% of genes in the cell.
Then, 10\% of genes are swapped within the cell (InnerSwap), before mutating 25\% of gene expression values with another random cell (CrossOver).
CLAIRE uses a neighborhood-based approach: mutual nearest neighbors (MNN) in the unintegrated space are computed for each cell across all batches. During augmentation, an inter- and an intra-batch views are computed by mutating between neighboring cells~\cite{XuhuaRFL23}. We also evaluate sampling positive pairs from a batch-balanced KNN (BBKNN) graph.
We investigate the impact of the MNN and BBKNN augmentations on the batch correction performance.

\paragraph{Domain-Specific Batch Normalization (DSBN):} Concerto~\cite{Concerto} adapts the idea of DSBN~\cite{DSBN}, a technique originally suggested for image data.
DSBN helps to learn domain-specific information to produce domain-invariant representations by applying separate batch normalization layers for each domain.
To investigate \emph{RQ3}, we replace the common batch normalization with DSBN where each experimental batch (different laboratory experiment in the same dataset) gets its own batch normalization layer, similar to Concerto.

\begin{table*}[t!]
\centering
\caption{Batch integration performance across five datasets. We show each method's biological conservation score (Bio), batch correction score (Batch), and total score (Total), with values computed across five runs with different random seeds. We group the methods by category (generic SSL, single-cell contrastive SSL frameworks, generative methods, and baselines). For uni-modal data (PBMC, Pancreas, and Immune Cell Atlas), the specialized encoder-decoder method scVI, the domain-specific SSL method CLAIRE, and a foundation model scGPT outperform other methods. For the multi-modal datasets PBMC-M and BMMC, generic methods achieve higher scores.}
\vspace{0.3cm}
\resizebox{\linewidth}{!}{
\setlength\tabcolsep{1.8pt}
\begin{tabular}{l|ccc|ccc||ccc|ccc|ccc}

\toprule
\multirow{2}{*}{\textbf{Method}} & \multicolumn{3}{c|}{\textbf{PBMC-M}} & \multicolumn{3}{c||}{\textbf{BMMC}} & \multicolumn{3}{c|}{\textbf{PBMC}} & \multicolumn{3}{c|}{\textbf{Pancreas}} & \multicolumn{3}{c}{\textbf{Immune Cell Atlas}} \\
& Bio & Batch & Total & Bio & Batch & Total & Bio & Batch & Total & Bio & Batch & Total & Bio & Batch & Total \\[-0.026cm]

\midrule

\multirow{2}{*}{SimCLR} & 0.877 & 0.434 & 0.700 & \textbf{0.877} & 0.601 & \textbf{0.767} & 0.370 & 0.563 & 0.447 & 0.791 & 0.615 & 0.721 & 0.555 & 0.753 & 0.635 \\[-0.026cm]
& \pmS{0.020}& \pmS{0.001}& \pmS{0.012}& \pmS{0.025}& \pmS{0.002}& \pmS{0.016} & \pmS{0.002} & \pmS{0.005} & \pmS{0.003} & \pmS{0.002} & \pmS{0.019} & \pmS{0.009} & \pmS{0.020} & \pmS{0.016} & \pmS{0.017} \\[-0.026cm]

\multirow{2}{*}{MoCo}    & 0.786 & \textbf{0.581} & 0.704 & 0.647 & \textbf{0.819} & 0.716 & 0.336 & 0.594 & 0.439 & 0.754 & 0.638 & 0.707 & 0.404 & \textbf{0.882} & 0.595 \\[-0.026cm]
& \pmS{0.005}& \pmS{0.016}& \pmS{0.003}& \pmS{0.048}& \pmS{0.024}& \pmS{0.038} & \pmS{0.007} & \pmS{0.014} & \pmS{0.010} & \pmS{0.008} & \pmS{0.017} & \pmS{0.011} & \pmS{0.024} & \pmS{0.022} & \pmS{0.016} \\[-0.026cm]

\multirow{2}{*}{SimSiam}& 0.903 & 0.455 & 0.724 & 0.753 & 0.571 & 0.680 & 0.271 & 0.512 & 0.368 & 0.531 & 0.635 & 0.572 & 0.358 & 0.640 & 0.470\\[-0.026cm]
& \pmS{0.057}& \pmS{0.029}& \pmS{0.046}& \pmS{0.007}& \pmS{0.002}& \pmS{0.005} & \pmS{0.016} & \pmS{0.002} & \pmS{0.011} & \pmS{0.113} & \pmS{0.017} & \pmS{0.061} & \pmS{0.040} & \pmS{0.020} & \pmS{0.030}\\[-0.026cm]                    
\multirow{2}{*}{NNCLR}  & 0.877 & 0.534 & 0.740 & 0.819 & 0.580 & 0.723 & 0.345 & 0.544 & 0.424 & 0.701 & 0.579 & 0.652 & 0.430 & 0.665 & 0.524 \\[-0.026cm]
& \pmS{0.033}& \pmS{0.004}& \pmS{0.018}& \pmS{0.021}& \pmS{0.008}& \pmS{0.016} & \pmS{0.011} & \pmS{0.009} & \pmS{0.010} & \pmS{0.052} & \pmS{0.015} & \pmS{0.037}  & \pmS{0.028} & \pmS{0.007} & \pmS{0.017} \\[-0.026cm]

\multirow{2}{*}{BYOL}   & \textbf{0.928} & 0.493 & \textbf{0.754} & 0.742 & 0.693 & 0.722 & 0.134 & 0.748 & 0.379 & 0.578 & 0.659 & 0.610 & 0.222 & 0.864 & 0.479 \\[-0.026cm]
& \pmS{0.065}& \pmS{0.016}& \pmS{0.033}& \pmS{0.043}& \pmS{0.016}& \pmS{0.019} & \pmS{0.017} & \pmS{0.076} & \pmS{0.020} & \pmS{0.029} & \pmS{0.012} & \pmS{0.013} & \pmS{0.031} & \pmS{0.009} & \pmS{0.021} \\[-0.026cm]

\multirow{2}{*}{VICReg} & 0.814 & 0.405 & 0.651 & 0.832 & 0.656 & 0.761 & 0.412 & 0.607 & 0.490 & 0.811 & 0.617 & \textbf{0.733} & 0.529 & 0.816 & 0.644 \\[-0.026cm]
& \pmS{0.039}& \pmS{0.026}& \pmS{0.013}& \pmS{0.051}& \pmS{0.009}& \pmS{0.027} & \pmS{0.010} & \pmS{0.000} & \pmS{0.006} & \pmS{0.003} & \pmS{0.001} & \pmS{0.002} & \pmS{0.014} & \pmS{0.022} & \pmS{0.012}\\[-0.026cm]

Barlow                  & 0.902 & 0.430 & 0.713 & 0.859 & 0.612 & 0.760 & 0.341 & 0.523 & 0.414 & 0.694 & 0.580 & 0.648 & 0.535 & 0.734 & 0.614 \\[-0.026cm]
Twins                   & \pmS{0.048}& \pmS{0.014}& \pmS{0.034}& \pmS{0.018}& \pmS{0.011}& \pmS{0.006} & \pmS{0.010} & \pmS{0.005} & \pmS{0.004} & \pmS{0.011} & \pmS{0.010} & \pmS{0.034} & \pmS{0.012} & \pmS{0.020} \\[-0.026cm]
\midrule

\multirow{2}{*}{Concerto} & 0.785 & 0.422 & 0.64 & 0.524 & 0.661 & 0.579 & 0.055 & 0.566 & 0.260 & 0.102 & 0.367 & 0.208 & 0.426 & 0.810 & 0.580 \\[-0.026cm]
& \pmS{0.002}& \pmS{0.003}& \pmS{0.002}& \pmS{0.019}& \pmS{0.008}& \pmS{0.015} & \pmS{0.000} & \pmS{0.000} & \pmS{0.000} & \pmS{0.003} & \pmS{0.000} & \pmS{0.002}& \pmS{0.014} & \pmS{0.025} & \pmS{0.017}\\[-0.026cm]

\multirow{2}{*}{CLEAR} & \multirow{2}{*}{---} & \multirow{2}{*}{---} & \multirow{2}{*}{---} & \multirow{2}{*}{---} & \multirow{2}{*}{---} & \multirow{2}{*}{---}  & 0.580 & 0.209 & 0.432 & 0.698 & 0.249 & 0.518 & 0.775 & 0.327 & 0.596 \\[-0.026cm]
&  &  &  &  &  &  & \pmS{0.000} & \pmS{0.002} & \pmS{0.001} & \pmS{0.011} & \pmS{0.002} & \pmS{0.006} & \pmS{0.037} & \pmS{0.008} & \pmS{0.022} \\[-0.026cm]

\multirow{2}{*}{CLAIRE} & \multirow{2}{*}{---} & \multirow{2}{*}{---} & \multirow{2}{*}{---} & \multirow{2}{*}{---} & \multirow{2}{*}{---} & \multirow{2}{*}{---} & 0.714 & 0.866 & 0.774 & 0.582 & \textbf{0.959} & 0.732 & 0.548 & 0.527 & 0.539 \\[-0.026cm]
&  &  &  &  &   & & \pmS{0.009} & \pmS{0.005} & \pmS{0.008} & \pmS{0.003} & \pmS{0.014} & \pmS{0.004} & \pmS{0.018} & \pmS{0.011} & \pmS{0.011}\\[-0.026cm]

\multirow{2}{*}{scCLIP} & 0.643 & 0.402 & 0.546 & 0.638 & 0.194 & 0.460 & \multirow{2}{*}{---} & \multirow{2}{*}{---} & \multirow{2}{*}{---} & \multirow{2}{*}{---} & \multirow{2}{*}{---} & \multirow{2}{*}{---} & \multirow{2}{*}{---} & \multirow{2}{*}{---} & \multirow{2}{*}{---} \\[-0.026cm]
                        & \pmS{0.002}& \pmS{0.004}& \pmS{0.001}& \pmS{0.006}& \pmS{0.005}& \pmS{0.005}&  &  &  &  &  &  &  &  &  \\[-0.026cm]

\midrule

scGPT & \multirow{2}{*}{---} & \multirow{2}{*}{---} & \multirow{2}{*}{---} & \multirow{2}{*}{---} & \multirow{2}{*}{---} & \multirow{2}{*}{---} & 0.440 & 0.469 & 0.451 & 0.473 & 0.168 & 0.351 & 0.380 & 0.516 & 0.435 \\[-0.026cm]
(zero-shot) &  &  &  &  &  &  & \pmS{0.010} & \pmS{0.017} & \pmS{0.013} & \pmS{0.001} & \pmS{0.003} & \pmS{0.002} & \pmS{0.012} & \pmS{0.014} & \pmS{0.008}\\[-0.026cm]

scGPT & \multirow{2}{*}{---} & \multirow{2}{*}{---} & \multirow{2}{*}{---} & \multirow{2}{*}{---} & \multirow{2}{*}{---} & \multirow{2}{*}{---} & \textbf{0.940} & 0.514 & 0.770 & \textbf{0.873} & 0.345 & 0.662 & \textbf{0.979} & 0.485 & \textbf{0.781} \\[-0.026cm]
(finetuned) &  &  &  &  &  &  & \pmS{0.012} & \pmS{0.011} & \pmS{0.011} & \pmS{0.003} & \pmS{0.047} & \pmS{0.021} & \pmS{0.017} & \pmS{0.019} & \pmS{0.017}\\[-0.026cm]

Geneformer & \multirow{2}{*}{---} & \multirow{2}{*}{---} & \multirow{2}{*}{---} & \multirow{2}{*}{---} & \multirow{2}{*}{---} & \multirow{2}{*}{---} & 0.024 & 0.462 & 0.199 & 0.004 & 0.437 & 0.177 & 0.013 & 0.265 & 0.114 \\[-0.026cm]
(finetuned) &  &  &  &  &  &  & \pmS{0.000} & \pmS{0.000} & \pmS{0.000} & \pmS{0.000} & \pmS{0.000} & \pmS{0.000} & \pmS{0.000} & \pmS{0.000} & \pmS{0.000}\\[-0.026cm]

\multirow{2}{*}{scButterfly} & 0.702 & 0.391 & 0.577 & 0.781 & 0.297 & 0.587 & \multirow{2}{*}{---} & \multirow{2}{*}{---} & \multirow{2}{*}{---} & \multirow{2}{*}{---} & \multirow{2}{*}{---} & \multirow{2}{*}{---} & \multirow{2}{*}{---} & \multirow{2}{*}{---} & \multirow{2}{*}{---} \\[-0.026cm]
& \pmS{0.002}& \pmS{0.003}& \pmS{0.002}& \pmS{0.000}& \pmS{0.004}& \pmS{0.002}&  &  &  &  &  &  &  &  &  \\[-0.026cm]

\multirow{2}{*}{scTEL} & 0.089 & 0.800 & 0.373 & 0.000 & 0.706 & 0.282  & \multirow{2}{*}{---} & \multirow{2}{*}{---} & \multirow{2}{*}{---} & \multirow{2}{*}{---} & \multirow{2}{*}{---} & \multirow{2}{*}{---} & \multirow{2}{*}{---} & \multirow{2}{*}{---} & \multirow{2}{*}{---} \\[-0.026cm]
& \pmS{0.002}& \pmS{0.004}& \pmS{0.001}& \pmS{0.006}& \pmS{0.005}& \pmS{0.005}&  &  &  &  &  &  &  &  &  \\[-0.026cm]

totalVI / & 0.702 & 0.305 & 0.543 & 0.755 & 0.272 & 0.562 & 0.918 & \textbf{0.871} & \textbf{0.899} & 0.805 & 0.511 & 0.688 & 0.862 & 0.593 & 0.754 \\[-0.026cm]
scVI & \pmS{0.002}& \pmS{0.002}& \pmS{0.001}& \pmS{0.002}& \pmS{0.001}& \pmS{0.002} & \pmS{0.015} & \pmS{0.000} & \pmS{0.009} & \pmS{0.002} & \pmS{0.007} & \pmS{0.001} & \pmS{0.033} & \pmS{0.013} & \pmS{0.024}  \\[-0.026cm]
 
\midrule
\multirow{2}{*}{SCDC} & \multirow{2}{*}{---} & \multirow{2}{*}{---} & \multirow{2}{*}{---} & \multirow{2}{*}{---} & \multirow{2}{*}{---} & \multirow{2}{*}{---} & 0.679 & 0.605 & 0.649 & 0.648 & 0.350 & 0.529 & 0.698 & 0.565 & 0.645 \\[-0.026cm]
&  &  &  &  &  &  & \pmS{0.050} & \pmS{0.005} & \pmS{0.028} & \pmS{0.018} & \pmS{0.023} & \pmS{0.002} & \pmS{0.024} & \pmS{0.019} & \pmS{0.020}\\[-0.026cm]

\multirow{2}{*}{PCA} & 0.448 & 0.369 & 0.417 & 0.538 & 0.282 & 0.436 & 0.558 & 0.303 & 0.456 & 0.683 & 0.266 & 0.516 & 0.677 & 0.276 & 0.517 \\[-0.026cm]
& \pmS{0.000}& \pmS{0.001}& \pmS{0.000}& \pmS{0.002}& \pmS{0.007}& \pmS{0.002}&\pmS{0.003} & \pmS{0.000} & \pmS{0.002} & \pmS{0.001} & \pmS{0.000} & \pmS{0.001} & \pmS{0.000} & \pmS{0.000} & \pmS{0.000}  \\[-0.026cm]
\bottomrule
\end{tabular}
}
\label{tab:clear_bc}
\vspace{-0.35cm}
\end{table*}

\paragraph{Multi-Modal Integration:}
For the multi-omics datasets PBMC-M and BMMC, we evaluate three integration methods as part of \emph{RQ3}.
First, addition takes two embeddings of the same dimensionality (one per modality) and adds them together to get a joint representation, similar to the Concerto~\cite{Concerto}.
Second, concatenation appends two embeddings. 
Third, instead of contrasting joint views of a cell, two modalities of the same cell are contrasted using a symmetric cross-entropy loss~\cite{wang2019symmetriccrossentropyrobust} and the CLIP approach.
After training with the CLIP approach~\cite{CLIP, scCLIP}, we concatenate two embeddings during inference.

\section{Experiments}
\label{sec:experiments}

As detailed in \autoref{sec:benchmark-design}, we benchmark nineteen SSL methods on nine single-cell datasets derived from different tissues with considerable variation in data size and complexity. 
See~\autoref{datasets} for details about the datasets.
All models are trained with five unique random seeds and we report their mean performance and standard deviation.

\subsection{Generic versus Specialized SSL Methods}
\label{sec:experiments_tasks}

\emph{RQ1} focuses on the comparison of specialized single-cell SSL frameworks and generic SSL methods. For that, we evaluate several models and two baselines on three important downstream tasks for uni- and multi-omics datasets.

\paragraph{Batch Correction:}
The batch correction performance of all methods across five datasets is presented in~\autoref{tab:clear_bc}. Our analysis includes two multi-modal (CITE-seq) datasets (PBMC-M and BMMC) and three single-modality (scRNA-seq) datasets (PBMC, Pancreas, and Immune Cell Atlas). 

For scRNA-seq datasets, our results show that scVI is the best-performing method that balances both batch correction and bio conservation (\autoref{tab:clear_bc}).
scVI performance drops for the MCA and Lung datasets (\autoref{tab:integration_supplement}) but excels for the Tabula Sapiens dataset (\autoref{tab:big_sapiens_bc}). CLAIRE ranks second-best overall but tends to overcorrect batch effects at the expense of biological variance. 
Overall and comparing to other single-cell generative models, finetuned scGPT performs well for the second-largest evaluated dataset, Immune Cell Atlas, by scoring high in bio conservation and total, but for smaller datasets batch score is significantly lower than other methods.
Zero-shot scGPT and finetuned Geneformer show unsatisfactory performance.
Across all common benchmarked SSL methods, VICReg, SimCLR, and MoCo perform satisfactorily for the Pancreas dataset. However, for the PBMC and Immune Cell Atlas datasets, these SSL methods prioritize batch correction over bio conservation as indicated by their high batch and low bio score. 
In comparison to the other methods, in all cases, Concerto significantly underperforms and achieves a total score lower than the baselines, PCA and SCDC.
As expected, PCA shows an adequate bio conservation score since it uses raw data and captures the true biological signal.

For multi-modal datasets, PBMC-M and BMMC, we observe that generic methods such as SimCLR, BYOL, MoCo, and VICReg are the best-performing methods, within their category and overall (\autoref{tab:clear_bc}).
Interestingly, MoCo overcorrects for all single- and multi-modal datasets.
The results show that there is room for improvement in specialized methods. Concerto, scCLIP, and scButterfly reach low batch correction results compared to general methods as MoCo or SimCLR. For PBMC-M, Concerto preserves biological variance and shows a high bio score.
scTEL succeeds at batch correction while failing in bio conservation and scoring almost zero for both multi-modal datasets.
Despite the success of scVI for uni-modal data, totalVI's results are unsatisfactory compared to the generic contrastive methods. For BMMC dataset, totalVI total score (0.562) is almost twice lower than then score of best method SimCLR (0.767).

\begin{figure}[t!]
  \centering
  \includegraphics[width=\linewidth]{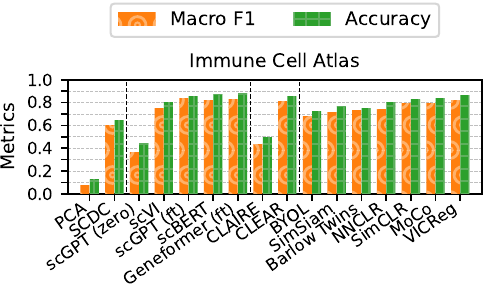}

  \vspace{-0.3cm}
  \caption{Uni-modal cell-typing with one sequencing technology (10X 5' v2) of the Immune Cell Atlas as a hold-out set. We train the encoder and classifier. The finetuned scGPT and Geneformer perform the best, while the generic VICReg method is a close third. The methods are grouped by category (baselines, specialized generative, specialized contrastive, and generic).}
  \label{fig:qr_big_transposed}
\end{figure}

\paragraph{Cell Type Annotation:}
We assess the cell-typing performance on the single-modal scRNA-seq datasets and CLEAR augmentations for the hold-out batch of the Immune Cell Atlas dataset (\autoref{fig:qr_big_transposed}) and all study data (\autoref{tab:template_qr}).
For all experiments, we do not use a projector during inference.
\autoref{tab:template_qr} evaluates cell-typing for the Pancreas dataset where unique batches were used as hold-out data.
The best-performing methods are VICReg, CLEAR, and in rare cases finetuned single-cell foundation models (scFMs). 
All generic SSL methods perform well, together with scVI, which takes additional information as input.
Although finetuned scFMs achieve adequate accuracy in most cases, they perform the best only for two datasets, the Immune Cell Atlas and the Xin study from the Pancreas, which are larger than the other three evaluated datasets. 
CLAIRE shows competitive results for the Pancreas dataset but falters in the second-biggest benchmarked dataset Immune Cell Atlas.
Additionally, CLAIRE and scFMs have a significantly higher computational load than the other methods.
PCA and zero-shot scGPT perform unsatisfactorily. 

\autoref{tab:multimodal_qr} shows cell-typing performance for multi-modal embeddings, where we integrate modalities through concatenation and train with CLEAR augmentations. 
We evaluate models using either both modalities (e.g., RNA and protein) or just the main modality (e.g., RNA) to assess whether representations capture information about the second modality and if a single main modality is sufficient during inference when the second modality is unavailable.
scButterfly slightly outperforms VICReg and SimCLR, that show competitive results, with scButterfly leading overall performance (\autoref{tab:multimodal_qr}).
All generic contrastive  methods achieve good accuracy and outperform specialized contrastive methods like Concerto and scCLIP.
While totalVI struggles with batch correction, it performs well in cell-typing.
All models except scCLIP show better performance with multi-omics data than single modality, though the performance drop for single-modality inference is minimal. Notably, scCLIP appears to treat the second modality as noise. We conclude that scButterfly and generic contrastive models can be used for a single modality inference if the second modality is missing. Of note, Concerto and totalVI do not support uni-modal inference for multi-modal data.

\begin{figure}[h]
  \centering
  \includegraphics[width=\linewidth]{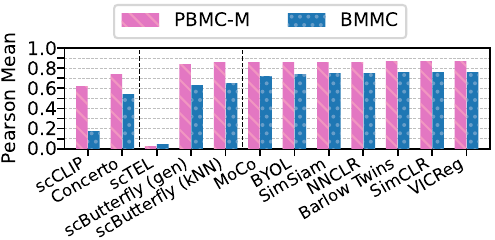}

  \vspace{-0.3cm}
  \caption{Missing modality prediction for models trained on the multi-modal datasets, PBMC and BMMC. We show the average Pearson correlation between the original and inferred missing modality: protein for PBMC-M and ADT (protein abundance) for BMMC. The methods are sorted from worst (left) to best (right) within group (specialized contrastive, generative, and generic).}
  \label{fig:multimodal_mp}
\end{figure}

\paragraph{Missing Modality Prediction:}
\autoref{fig:multimodal_mp} shows the ability to predict missing protein values while given only RNA or gene expression (GEX) during inference. 
The model is trained on multi-omics data using CLEAR augmentations and concatenation to combine modalities. The standard deviation is close to zero, see~\autoref{tab:multimodal_mp}. 
VICReg and SimCLR outperform other methods, including specialized single-cell frameworks.
We assume that Concerto, scCLIP, and scTEL do not learn enough information about the secondary modality (protein) and its connection to the main modality (RNA) and, therefore, are not able to predict the missing modality.
We evaluate scButterfly in two modes: averaging kNN, as done for other methods, and generating proteins directly from gene expression data. While the performance differences between the two approaches are insignificant, the scButterfly's Pearson correlation is unsatisfactory low compared to generic contrastive methods.
High values of Pearson's correlations show that models effectively infer protein values from gene expression data.

\paragraph{Summary and Findings:} Overall, specialized SSL methods designed for single-cell analysis have not demonstrated clear advantages over general-purpose approaches, except for scVI, finetuned scGPT, and CLAIRE. We attribute the superior performance of scVI, finetuned scGPT, and CLAIRE to the fact that these methods leverage experimental batch information during training.  
For multi-omics data, the best models are SimCLR and VICReg. Our findings indicate that current single-cell SSL methods such as scCLIP or totalVI need improvement for multi-modal downstream tasks, as they do not yet surpass or compete with generic architectures in performance.

\subsection{Ablation Study}
\label{sec:experiments_hyperparam}

\emph{RQ2} investigates how hyperparameters and augmentations impact the performance of single-cell SSL. We conduct hyperparameter tuning for all generic methods using two datasets: HIC and MCA. For the mentioned frameworks, we use the hyperparameters proposed in the respective original papers. For the generic methods, we focus on augmentations, the representation dimensionality, the projection dimensionality, and the temperature parameter.  

\paragraph{Representation Dimensionality:} We perform a grid search over the representation dimensionality for the HIC and MCA datasets, evaluating the batch correction performance (details in~\autoref{sec:hparams}). We train all models with embedding dimensions \{8, 16, 32, 64, 128, 256, 512, 1024\}. Models are ranked according to the \textsc{scib}~\cite{luecken2022benchmarking} total score, which is min-max scaled across all models. Our findings indicate that lower dimensionalities of 64 and 128 consistently lead to the best performance across all considered methods, while the larger dimensionality of 1024 achieves similar but requires more training time and memory (\autoref{fig:rep-dim} in Appendix).
Given these observations, we adopt the embedding size of 64 for subsequent experiments.

\paragraph{Projector Dimensionality:}
To learn more robust representations, self-supervised models may benefit from projection heads~\cite{xue2024investigatingbenefitsprojectionhead}. 
We investigate the impact of projection dimensionality during training by introducing a scale factor. 
At inference time, the projection head is discarded, and only the encoder is used. 
For contrastive methods, the projection size is scaled down by this factor, while for non-contrastive methods, it is scaled up by the same factor (see~\autoref{projection-def}). 
Our results reveal that the projector effect is ambiguous for most models (\autoref{fig:projection-dim}). 
However, BarlowTwins, BYOL, and VICReg show an improved performance with larger scaling factors.

\begin{figure}[t!]
  \centering
  
  \includegraphics[width=\linewidth]{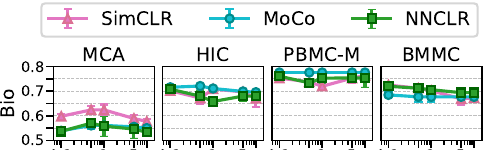}

  \vspace{0.03cm}
  
  \includegraphics[width=\linewidth]{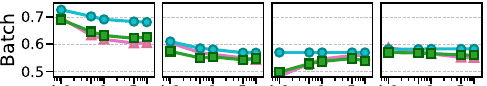}
  
  \vspace{0.03cm}
  
  \includegraphics[width=\linewidth]{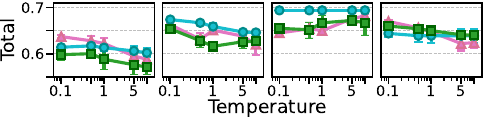}

  \vspace{-0.3cm}
  \caption{Temperature impact on the loss of three contrastive methods on four datasets (columns). Bio conservation, batch correction, and total scores are represented on the y-axis. The results are not min-max scaled for easier comparison. Overall, smaller temperature leads to better data integration.}
  \label{fig:temp-ablations}
\end{figure}

\paragraph{Temperature Impact:} 
\autoref{fig:temp-ablations} shows the temperature $t^{\circ}$ effect for SimCLR, MoCo, and NNCLR models. We evaluate $t^{\circ} \in \{0.1, 0.5, 1, 5, 10\}$ using \textsc{scib-metrics} scores.
Overall, lower $t^{\circ}$ values lead to better scores.
The PyTorch default value ($t^{\circ}=0.5$) performs well across all models and datasets (used for subsequent experiments).
For scRNA-seq datasets, MCA and HIC, batch correction metric decreases with increasing $t^{\circ}$, except for NNCLR model and HIC dataset.
For HIC data, higher $t^{\circ}$ leads to uncorrected batch effects, while
for MCA data, it results in clustering of similar cell types that are expected to be near but not fully mixed together (e.g., CD4 T and CD8 T cells).
For multi-modal datasets, PBMC-M and BMMC, $t^{\circ}$ effects are inconsistent.
For PBMC-M, MoCo produces extremely similar results with all evaluated $t^{\circ}$, while SimCLR and NNCLR mix batches better with higher $t^{\circ}$, e.g., with a $t^{\circ}$ of 10 cell types are separated with mixed batches.
For BMMC, lower $t^{\circ}$ achieves better batch correction.

\begin{figure}[t!]
  \centering
  \includegraphics[width=\linewidth]{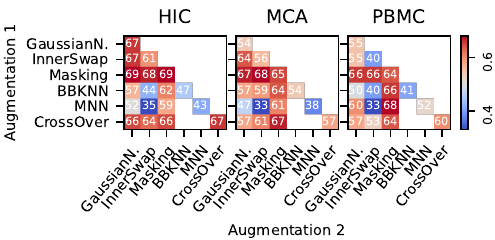
  }

  \vspace{-0.3cm}
      
  \caption{Evaluation of individual and combined data augmentations for the VICReg method based on total score for batch correction. Diagonal entries correspond to a single augmentation, and off-diagonal entries correspond to the two sequentially applied augmentations. Hyperparameters are based on ablation results~(\autoref{tab:augmentation-parameters}), evaluation for SimCLR and MoCo are in \autoref{fig:heatmaps_aug_more}.}
  \label{fig:heatmaps}

  \vspace{-0.1cm}
\end{figure}

\paragraph{Augmentation Ablation:} 
The space of augmentations in the single-cell domain can be split into: random transformations~\cite{CLEAR,Richter2024.02.16.580624} and neighborhood-based transformations~\cite{XuhuaRFL23, liu2024cake}. 
We perform an ablation for all studied augmentations and optimize hyperparameters for each (see~\autoref{sec:aug-ablation}).
To study how augmentations affect each other, we train VICReg, SimCLR, and MoCo models with combinations of two augmentations. 
We choose these models due to their consistently good performance.
Random masking is the best-performing augmentation alone and combined with others (\autoref{fig:heatmaps}).
Additionally, CrossOver performs competitively, especially for the SimCLR model (\autoref{fig:heatmaps_aug_more}).


\paragraph{Summary and Findings:} Higher-dimensional embeddings and lower temperatures enhance consistency and performance. Larger representations have better bio conservation, smaller - batch correction. However, the optimal embedding size is 64 or 128. Among various data augmentation techniques, masking proves most effective and surpasses even sophisticated biology-specific approaches that incorporate batch information, such as MNN and BBKNN.

\subsection{Impact of Batch Normalization and Multi-Modal Integration Proposed for Image Data}
\label{sec:experiments:common_com_practices}

In the following, we address \emph{RQ3} and study the impact of common SSL techniques, such as retaining the projector and domain-specific batch normalization~\cite{DSBN}.

\paragraph{Retaining Projector:}
To evaluate the impact of the projector during inference, we train a model consisting of an encoder and projector with CLEAR augmentations, and evaluate it with/without projection.
Although it is common to only use the encoder during inference to create an embedding~\cite{chen2020simple, chen2020exploringsimplesiameserepresentation}, the projector is also leveraged in the single-cell community and we compare the two approaches. 
First, we analyze how retaining the projection layer affects batch integration for single-modal data (\autoref{tab:unimodal_projection}). 
While using the projection layer slightly improves batch correction, it generally reduces biological conservation and total (which weights bio conservation more) scores. 
For example, the total score decreases from 0.625 to 0.608 for SimCLR on the MCA dataset. These findings suggest using only the encoder during inference, rather than the combined encoder and projector. The former better preserves biological signals despite slightly worse batch correction performance.
Second, we observe a similar trend on multi-modal datasets as on scRNA-seq data (\autoref{tab:multimodal_projection_bc}).
The effects are less consistent and conclusive (scores change among models/datasets), and only MoCo's batch correction benefits from projection. 
Remarkably, Concerto uses encoder and projector during inference but has comparably unsatisfactory batch correction performance.

\paragraph{Domain-Specific Batch Normalization (DSBN):} 
Inspired by Concerto~\cite{Concerto} and common practices from computer vision applications, we evaluate whether models benefit from DSBN~\cite{DSBN}.
Although the Concerto~\cite{Concerto} manuscript discusses the usage of DSBN, the publicly available code does not apply DSBN. 
Therefore, we evaluate DSBN only for generic methods.
\autoref{tab:unimodal_dsbn} shows reduced total performance when leveraging DSBN compared to standard batch normalization.
For HIC dataset, DSBN leads to slightly better batch correction but worse bio conservation.
However, it is not the case for the MCA dataset.

\paragraph{Multi-Modal Integration Methods:}
In~\autoref{tab:Neurips_compare_add_concat_clip_bc}, we compare three ways to combine multiple modalities of a cell: element-wise addition of uni-modal embeddings~\cite{Concerto}, concatenation of uni-modal embeddings, and multi-modal contrastive learning with the CLIP objective~\cite{CLIP, scCLIP}.
For each modality, we train a model with CLEAR~\cite{CLEAR} augmentations
and discard the projector during inference.
See~\autoref{sec:multimodal_setting} for details.
\autoref{tab:Neurips_compare_add_concat_clip_bc} shows that concatenation is the best integration method.
Addition and concatenation show high results in bio conservation, while the CLIP-based approach performs better in batch correction.

\paragraph{Summary and Findings:} 
Previously recommended techniques, such as keeping the projector or using DSBN, fail to enhance performance. 
For combining multiple modalities, concatenation turns out to be the most effective approach.

\section{Conclusions}
\label{sec:conclusions}
We introduced a comprehensive benchmark, scSSL-Bench, for self-supervised learning on uni- and multi-modal single-cell data.
First, we observe that specialized single-cell SSL methods perform better than generic methods for uni-modal data and underperform for multi-modal data (\emph{RQ1}). The best scRNA-seq single-modal data integration methods are scVI, CLAIRE, and the finetuned scGPT, all specialized for single-cell data.
Generative scVI and the single-cell foundation model scGPT significantly outperform all SSL methods, while CLAIRE shows good scores only for a subset of the datasets.
For multi-omics data, the generic methods SimCLR and VICReg perform the best and even outperform all other methods in the cell type annotation and missing modality prediction tasks for single-modal data.
According to our findings, there is a need for improving existing and developing new multi-modal specialized SSL methods since current existing frameworks do not outperform generic architectures, and multi-modal integration turned out to be a more difficult task than uni-modal (\emph{RQ1}).
Second, we conclude that masking augmentation leads to the biggest improvements alone and in combination with other types of augmentations, and moderately-size to large embedding sizes lead to better results (\emph{RQ2}).
Third, retaining the projection head or applying domain-specific batch normalization do not significantly influence the scores and rather degrade the total data integration score by achieving a lower bio conservation and higher batch correction indicating primarily regularization potential (\emph{RQ3}).
Finally, our benchmark offers a standardized framework for assessing new SSL methods, enabling researchers to systematically evaluate and compare their approaches against established baselines.

\section*{Acknowledgements}

The authors thank Sebastian Baunsgaard, Dmitry Kobak, and Thomas Sutter for their insightful comments and constructive feedback on the manuscript. We thank the three anonymous reviewers for their thorough reviews, which significantly improved the quality of this paper.

Computational data analysis was performed at Leonhard Med secure trusted research environment at ETH Zurich and  at the BIFOLD Hydra cluster.

FB is supported by the Swiss National Science Foundation (SNSF) (grant number 205321-207931).



\section*{Impact Statement}
This work provides a systematic benchmark of self-supervised learning (SSL) methods in single-cell genomics; we evaluate nineteen approaches across batch correction, cell typing, and missing modality prediction tasks. Our results offer practical guidelines for applying SSL to biological data, advancing computational tools for single-cell analysis.
The implications span from biomedical research to personalized medicine, where improved data integration enables better diagnostic and therapeutic strategies. By standardizing evaluation protocols, this benchmark promotes reproducibility and cross-disciplinary collaboration.

\balance
\bibliography{refs}

\begin{thebibliography}{72}
\providecommand{\natexlab}[1]{#1}
\providecommand{\url}[1]{\texttt{#1}}
\expandafter\ifx\csname urlstyle\endcsname\relax
  \providecommand{\doi}[1]{doi: #1}\else
  \providecommand{\doi}{doi: \begingroup \urlstyle{rm}\Url}\fi

\bibitem[Bardes et~al.(2022)Bardes, Ponce, and LeCun]{vicreg}
Bardes, A., Ponce, J., and LeCun, Y.
\newblock {VICReg: Variance-Invariance-Covariance Regularization for Self-Supervised Learning}.
\newblock In \emph{ICLR}, 2022.

\bibitem[Baysoy et~al.(2023)Baysoy, Bai, Satija, and Fan]{Baysoy2023}
Baysoy, A., Bai, Z., Satija, R., and Fan, R.
\newblock {The technological landscape and applications of single-cell multi-omics}.
\newblock \emph{Nature Reviews Molecular Cell Biology}, 24\penalty0 (10):\penalty0 695--713, Oct 2023.
\newblock ISSN 1471-0080.
\newblock \doi{10.1038/s41580-023-00615-w}.
\newblock URL \url{https://doi.org/10.1038/s41580-023-00615-w}.

\bibitem[B{\"u}ttner et~al.(2019)B{\"u}ttner, Miao, Wolf, Teichmann, and Theis]{Buttner2019}
B{\"u}ttner, M., Miao, Z., Wolf, F.~A., Teichmann, S.~A., and Theis, F.~J.
\newblock {A test metric for assessing single-cell RNA-seq batch correction}.
\newblock \emph{Nature Methods}, 16\penalty0 (1):\penalty0 43--49, Jan 2019.
\newblock ISSN 1548-7105.
\newblock \doi{10.1038/s41592-018-0254-1}.
\newblock URL \url{https://doi.org/10.1038/s41592-018-0254-1}.

\bibitem[Cao et~al.(2024{\natexlab{a}})Cao, Nai, Yang, Huang, and Gao]{CaoNegativeFree}
Cao, J., Nai, R., Yang, Q., Huang, J., and Gao, Y.
\newblock {An empirical study on disentanglement of negative-free contrastive learning}.
\newblock In \emph{NeurIPS}, 2024{\natexlab{a}}.
\newblock ISBN 9781713871088.

\bibitem[Cao et~al.(2024{\natexlab{b}})Cao, Zhao, Tang, Jiang, Li, Li, and Chen]{scButterfly}
Cao, Y., Zhao, X., Tang, S., Jiang, Q., Li, S., Li, S., and Chen, S.
\newblock {scButterfly: a versatile single-cell cross-modality translation method via dual-aligned variational autoencoders}.
\newblock \emph{{Nature Communications}}, 15\penalty0 (1):\penalty0 2973, Apr 2024{\natexlab{b}}.
\newblock ISSN 2041-1723.
\newblock \doi{10.1038/s41467-024-47418-x}.
\newblock URL \url{https://doi.org/10.1038/s41467-024-47418-x}.

\bibitem[Chang et~al.(2019)Chang, You, Seo, Kwak, and Han]{DSBN}
Chang, W.-G., You, T., Seo, S., Kwak, S., and Han, B.
\newblock Domain-specific batch normalization for unsupervised domain adaptation.
\newblock In \emph{Proceedings of the IEEE/CVF conference on Computer Vision and Pattern Recognition}, pp.\  7354--7362, 2019.

\bibitem[Chen et~al.(2020)Chen, Kornblith, Norouzi, and Hinton]{chen2020simple}
Chen, T., Kornblith, S., Norouzi, M., and Hinton, G.
\newblock {A simple framework for contrastive learning of visual representations}.
\newblock In \emph{ICML}, pp.\  1597--1607, 2020.

\bibitem[Chen \& He(2020)Chen and He]{chen2020exploringsimplesiameserepresentation}
Chen, X. and He, K.
\newblock {Exploring Simple Siamese Representation Learning}, 2020.
\newblock URL \url{https://arxiv.org/abs/2011.10566}.

\bibitem[Chen et~al.(2025)Chen, Fan, Shi, Shi, and Wang]{scTEL}
Chen, Y., Fan, X., Shi, C., Shi, Z., and Wang, C.
\newblock {A joint analysis of single cell transcriptomics and proteomics using transformer}.
\newblock \emph{npj Systems Biology and Applications}, 11\penalty0 (1):\penalty0 1, Jan 2025.
\newblock ISSN 2056-7189.
\newblock \doi{10.1038/s41540-024-00484-9}.
\newblock URL \url{https://doi.org/10.1038/s41540-024-00484-9}.

\bibitem[Conde et~al.(2022)Conde, Xu, Jarvis, Rainbow, Wells, et~al.]{doi:10.1126/science.abl5197}
Conde, C.~D., Xu, C., Jarvis, L.~B., Rainbow, D.~B., Wells, S.~B., et~al.
\newblock Cross-tissue immune cell analysis reveals tissue-specific features in humans.
\newblock \emph{Science}, 376\penalty0 (6594):\penalty0 eabl5197, 2022.
\newblock \doi{10.1126/science.abl5197}.
\newblock URL \url{https://www.science.org/doi/abs/10.1126/science.abl5197}.

\bibitem[Cui et~al.(2024)Cui, Wang, Maan, Pang, Luo, Duan, and Wang]{cui2024scgpt}
Cui, H., Wang, C., Maan, H., Pang, K., Luo, F., Duan, N., and Wang, B.
\newblock {scGPT: toward building a foundation model for single-cell multi-omics using generative AI}.
\newblock \emph{Nature Methods}, 21\penalty0 (8):\penalty0 1470--1480, 2024.

\bibitem[Devlin et~al.(2019{\natexlab{a}})Devlin, Chang, Lee, and Toutanova]{devlin-etal-2019-bert}
Devlin, J., Chang, M.-W., Lee, K., and Toutanova, K.
\newblock {BERT: Pre-training of Deep Bidirectional Transformers for Language Understanding}.
\newblock In Burstein, J., Doran, C., and Solorio, T. (eds.), \emph{Proceedings of the 2019 Conference of the North {A}merican Chapter of the Association for Computational Linguistics: Human Language Technologies, Volume 1 (Long and Short Papers)}, pp.\  4171--4186, Minneapolis, Minnesota, June 2019{\natexlab{a}}. Association for Computational Linguistics.
\newblock \doi{10.18653/v1/N19-1423}.
\newblock URL \url{https://aclanthology.org/N19-1423/}.

\bibitem[Devlin et~al.(2019{\natexlab{b}})Devlin, Chang, Lee, and Toutanova]{devlin2019bert}
Devlin, J., Chang, M.-W., Lee, K., and Toutanova, K.
\newblock {Bert: Pre-training of deep bidirectional transformers for language understanding}.
\newblock In \emph{Proceedings of the 2019 conference of the North American chapter of the association for computational linguistics: human language technologies, volume 1 (long and short papers)}, pp.\  4171--4186, 2019{\natexlab{b}}.

\bibitem[Ding et~al.(2020)Ding, Adiconis, Simmons, Kowalczyk, Hession, et~al.]{Ding632216}
Ding, J., Adiconis, X., Simmons, S.~K., Kowalczyk, M.~S., Hession, C.~C., et~al.
\newblock Systematic comparison of single-cell and single-nucleus rna-sequencing methods.
\newblock \emph{{Nature Biotechnology}}, 38\penalty0 (6):\penalty0 737--746, Jun 2020.
\newblock ISSN 1546-1696.
\newblock \doi{10.1038/s41587-020-0465-8}.
\newblock URL \url{https://doi.org/10.1038/s41587-020-0465-8}.

\bibitem[Dwibedi et~al.(2021)Dwibedi, Aytar, Tompson, Sermanet, and Zisserman]{dwibedi2021littlehelpfriendsnearestneighbor}
Dwibedi, D., Aytar, Y., Tompson, J., Sermanet, P., and Zisserman, A.
\newblock {With a Little Help from My Friends: Nearest-Neighbor Contrastive Learning of Visual Representations}, 2021.
\newblock URL \url{https://arxiv.org/abs/2104.14548}.

\bibitem[Eraslan et~al.(2022)Eraslan, Drokhlyansky, Anand, Fiskin, Subramanian, Slyper, Wang, Van~Wittenberghe, Rouhana, Waldman, et~al.]{eraslan2022single}
Eraslan, G., Drokhlyansky, E., Anand, S., Fiskin, E., Subramanian, A., Slyper, M., Wang, J., Van~Wittenberghe, N., Rouhana, J.~M., Waldman, J., et~al.
\newblock {Single-nucleus cross-tissue molecular reference maps toward understanding disease gene function}.
\newblock \emph{Science}, 376\penalty0 (6594):\penalty0 eabl4290, 2022.

\bibitem[Gayoso et~al.(2021)Gayoso, Steier, Lopez, Regier, Nazor, Streets, and Yosef]{Gayoso2021}
Gayoso, A., Steier, Z., Lopez, R., Regier, J., Nazor, K.~L., Streets, A., and Yosef, N.
\newblock {Joint probabilistic modeling of single-cell multi-omic data with totalVI}.
\newblock \emph{Nature Methods}, 18\penalty0 (3):\penalty0 272--282, Mar 2021.
\newblock ISSN 1548-7105.
\newblock \doi{10.1038/s41592-020-01050-x}.
\newblock URL \url{https://doi.org/10.1038/s41592-020-01050-x}.

\bibitem[Geiping et~al.(2023)Geiping, Garrido, Fernandez, Bar, Pirsiavash, LeCun, and Goldblum]{geiping2023cookbook}
Geiping, J., Garrido, Q., Fernandez, P., Bar, A., Pirsiavash, H., LeCun, Y., and Goldblum, M.
\newblock {A Cookbook of Self-Supervised Learning}, 2023.
\newblock URL \url{https://arxiv.org/abs/2304.12210}.

\bibitem[Grandi et~al.(2022)Grandi, Modi, Kampman, and Corces]{ATAC-seq}
Grandi, F.~C., Modi, H., Kampman, L., and Corces, M.~R.
\newblock {Chromatin accessibility profiling by ATAC-seq}.
\newblock \emph{Nature Protocols}, 17\penalty0 (6):\penalty0 1518--1552, Jun 2022.
\newblock ISSN 1750-2799.
\newblock \doi{10.1038/s41596-022-00692-9}.
\newblock URL \url{https://doi.org/10.1038/s41596-022-00692-9}.

\bibitem[Grill et~al.(2020)Grill, Strub, Altch{\'e}, Tallec, Richemond, Buchatskaya, Doersch, Avila~Pires, Guo, Gheshlaghi~Azar, et~al.]{grill2020bootstrap}
Grill, J.-B., Strub, F., Altch{\'e}, F., Tallec, C., Richemond, P., Buchatskaya, E., Doersch, C., Avila~Pires, B., Guo, Z., Gheshlaghi~Azar, M., et~al.
\newblock Bootstrap your own latent-a new approach to self-supervised learning.
\newblock \emph{NeurIPS}, 33:\penalty0 21271--21284, 2020.

\bibitem[Han et~al.(2022)Han, Cheng, Chen, Zhong, Hu, Chen, Zong, Hong, Chan, King, Gao, and Li]{CLEAR}
Han, W., Cheng, Y., Chen, J., Zhong, H., Hu, Z., Chen, S., Zong, L., Hong, L., Chan, T.-F., King, I., Gao, X., and Li, Y.
\newblock {Self-supervised contrastive learning for integrative single cell RNA-seq data analysis}.
\newblock \emph{Briefings in Bioinformatics}, 23\penalty0 (5):\penalty0 bbac377, 09 2022.
\newblock ISSN 1477-4054.
\newblock \doi{10.1093/bib/bbac377}.

\bibitem[Hao et~al.(2021)Hao, Hao, Andersen-Nissen, Mauck, Zheng, Butler, Lee, et~al.]{HAO20213573}
Hao, Y., Hao, S., Andersen-Nissen, E., Mauck, W.~M., Zheng, S., Butler, A., Lee, M.~J., et~al.
\newblock Integrated analysis of multimodal single-cell data.
\newblock \emph{Cell}, 184\penalty0 (13):\penalty0 3573--3587.e29, 2021.
\newblock ISSN 0092-8674.
\newblock \doi{https://doi.org/10.1016/j.cell.2021.04.048}.
\newblock URL \url{https://www.sciencedirect.com/science/article/pii/S0092867421005833}.

\bibitem[Hao et~al.(2024)Hao, Stuart, Kowalski, Choudhary, Hoffman, et~al.]{seurat}
Hao, Y., Stuart, T., Kowalski, M.~H., Choudhary, S., Hoffman, P., et~al.
\newblock {Dictionary learning for integrative, multimodal and scalable single-cell analysis}.
\newblock \emph{{Nature Biotechnology}}, 42\penalty0 (2):\penalty0 293--304, Feb 2024.
\newblock ISSN 1546-1696.
\newblock \doi{10.1038/s41587-023-01767-y}.
\newblock URL \url{https://doi.org/10.1038/s41587-023-01767-y}.

\bibitem[He et~al.(2020)He, Fan, Wu, Xie, and Girshick]{he2020momentum}
He, K., Fan, H., Wu, Y., Xie, S., and Girshick, R.
\newblock {Momentum contrast for unsupervised visual representation learning}.
\newblock In \emph{Proceedings of the IEEE/CVF conference on computer vision and pattern recognition}, pp.\  9729--9738, 2020.

\bibitem[Heryanto et~al.(2024)Heryanto, Zhang, and Imoto]{Heryanto2024}
Heryanto, Y.~D., Zhang, Y.-z., and Imoto, S.
\newblock {Predicting cell types with supervised contrastive learning on cells and their types}.
\newblock \emph{Scientific Reports}, 14\penalty0 (1):\penalty0 430, Jan 2024.
\newblock ISSN 2045-2322.
\newblock \doi{10.1038/s41598-023-50185-2}.
\newblock URL \url{https://doi.org/10.1038/s41598-023-50185-2}.

\bibitem[Heumos et~al.(2023)Heumos, Schaar, Lance, Litinetskaya, Drost, Zappia, L{\"u}cken, Strobl, Henao, Curion, et~al.]{heumos2023best}
Heumos, L., Schaar, A.~C., Lance, C., Litinetskaya, A., Drost, F., Zappia, L., L{\"u}cken, M.~D., Strobl, D.~C., Henao, J., Curion, F., et~al.
\newblock Best practices for single-cell analysis across modalities.
\newblock \emph{Nature Reviews Genetics}, 24\penalty0 (8):\penalty0 550--572, 2023.

\bibitem[Hu et~al.(2023)Hu, Li, Liu, Wu, Chen, Wang, and Liu]{hu2023teacherstudentarchitectureknowledgedistillation}
Hu, C., Li, X., Liu, D., Wu, H., Chen, X., Wang, J., and Liu, X.
\newblock {Teacher-Student Architecture for Knowledge Distillation: A Survey}, 2023.
\newblock URL \url{https://arxiv.org/abs/2308.04268}.

\bibitem[Jing et~al.(2022)Jing, Vincent, LeCun, and Tian]{jing2022understanding}
Jing, L., Vincent, P., LeCun, Y., and Tian, Y.
\newblock {Understanding Dimensional Collapse in Contrastive Self-supervised Learning}.
\newblock In \emph{International Conference on Learning Representations}, 2022.
\newblock URL \url{https://openreview.net/forum?id=YevsQ05DEN7}.

\bibitem[Jones et~al.(2022)Jones, Karkanias, Krasnow, Pisco, Quake, et~al.]{tabula}
Jones, R.~C., Karkanias, J., Krasnow, M.~A., Pisco, A.~O., Quake, S.~R., et~al.
\newblock {The Tabula Sapiens: A multiple-organ, single-cell transcriptomic atlas of humans}.
\newblock \emph{Science}, 376\penalty0 (6594):\penalty0 eabl4896, 2022.
\newblock \doi{10.1126/science.abl4896}.
\newblock URL \url{https://www.science.org/doi/abs/10.1126/science.abl4896}.

\bibitem[Kingma \& Ba(2017)Kingma and Ba]{kingma2017adammethodstochasticoptimization}
Kingma, D.~P. and Ba, J.
\newblock {Adam: A Method for Stochastic Optimization}, 2017.
\newblock URL \url{https://arxiv.org/abs/1412.6980}.

\bibitem[L{\"a}hnemann et~al.(2020)L{\"a}hnemann, K{\"o}ster, Szczurek, McCarthy, Hicks, Robinson, Vallejos, Campbell, Beerenwinkel, Mahfouz, et~al.]{lahnemann2020eleven}
L{\"a}hnemann, D., K{\"o}ster, J., Szczurek, E., McCarthy, D.~J., Hicks, S.~C., Robinson, M.~D., Vallejos, C.~A., Campbell, K.~R., Beerenwinkel, N., Mahfouz, A., et~al.
\newblock {Eleven grand challenges in single-cell data science}.
\newblock \emph{Genome biology}, 21\penalty0 (1):\penalty0 1--35, 2020.

\bibitem[Lance et~al.(2022)Lance, Luecken, Burkhardt, Cannoodt, Rautenstrauch, Laddach, Ubingazhibov, Cao, Deng, Khan, Liu, Russkikh, et~al.]{NeuripsDataChallengePaper}
Lance, C., Luecken, M.~D., Burkhardt, D.~B., Cannoodt, R., Rautenstrauch, P., Laddach, A., Ubingazhibov, A., Cao, Z.-J., Deng, K., Khan, S., Liu, Q., Russkikh, N., et~al.
\newblock {Multimodal single cell data integration challenge: Results and lessons learned}.
\newblock In \emph{NeurIPS Competitions and Demonstrations Track}, volume 176, pp.\  162--176, 2022.
\newblock URL \url{https://proceedings.mlr.press/v176/lance22a.html}.

\bibitem[Li et~al.(2023)Li, Zhang, Yang, Peng, Yu, Liu, Lv, Chen, and Peng]{scBridge}
Li, Y., Zhang, D., Yang, M., Peng, D., Yu, J., Liu, Y., Lv, J., Chen, L., and Peng, X.
\newblock {scBridge embraces cell heterogeneity in single-cell RNA-seq and ATAC-seq data integration}.
\newblock \emph{Nature Communications}, 14\penalty0 (1):\penalty0 6045, Sep 2023.
\newblock ISSN 2041-1723.
\newblock \doi{10.1038/s41467-023-41795-5}.
\newblock URL \url{https://doi.org/10.1038/s41467-023-41795-5}.

\bibitem[Li et~al.(2024)Li, Lin, Hu, Peng, Luo, and Peng]{SCDC}
Li, Y., Lin, Y., Hu, P., Peng, D., Luo, H., and Peng, X.
\newblock {Single-Cell RNA-Seq Debiased Clustering via Batch Effect Disentanglement}.
\newblock \emph{IEEE Transactions on Neural Networks and Learning Systems}, 35\penalty0 (8):\penalty0 11371--11381, 2024.
\newblock \doi{10.1109/TNNLS.2023.3260003}.

\bibitem[Liu et~al.(2024)Liu, Zeng, Kan, Li, and Zheng]{liu2024cake}
Liu, J., Zeng, W., Kan, S., Li, M., and Zheng, R.
\newblock {CAKE: a flexible self-supervised framework for enhancing cell visualization, clustering and rare cell identification}.
\newblock \emph{Briefings in Bioinformatics}, 25\penalty0 (1):\penalty0 bbad475, 2024.

\bibitem[Lopez et~al.(2018)Lopez, Regier, Cole, Jordan, and Yosef]{scvi}
Lopez, R., Regier, J., Cole, M.~B., Jordan, M.~I., and Yosef, N.
\newblock {Deep generative modeling for single-cell transcriptomics}.
\newblock \emph{Nature Methods}, 15\penalty0 (12):\penalty0 1053--1058, Dec 2018.
\newblock ISSN 1548-7105.
\newblock \doi{10.1038/s41592-018-0229-2}.

\bibitem[Lotfollahi et~al.(2022)Lotfollahi, Naghipourfar, Luecken, Khajavi, B{\"u}ttner, Wagenstetter, Avsec, Gayoso, Yosef, Interlandi, Rybakov, Misharin, and Theis]{Lotfollahi2022}
Lotfollahi, M., Naghipourfar, M., Luecken, M.~D., Khajavi, M., B{\"u}ttner, M., Wagenstetter, M., Avsec, {\v{Z}}., Gayoso, A., Yosef, N., Interlandi, M., Rybakov, S., Misharin, A.~V., and Theis, F.~J.
\newblock {Mapping single-cell data to reference atlases by transfer learning}.
\newblock \emph{{Nature Biotechnology}}, 40\penalty0 (1):\penalty0 121--130, Jan 2022.
\newblock ISSN 1546-1696.
\newblock \doi{10.1038/s41587-021-01001-7}.
\newblock URL \url{https://doi.org/10.1038/s41587-021-01001-7}.

\bibitem[Luecken et~al.(2021)Luecken, Burkhardt, Cannoodt, Lance, Agrawal, Aliee, Chen, Deconinck, Detweiler, Granados, et~al.]{NeuripsDataChallenge}
Luecken, M., Burkhardt, D., Cannoodt, R., Lance, C., Agrawal, A., Aliee, H., Chen, A., Deconinck, L., Detweiler, A., Granados, A., et~al.
\newblock {A sandbox for prediction and integration of DNA, RNA, and proteins in single cells}.
\newblock In Vanschoren, J. and Yeung, S. (eds.), \emph{NeurIPS Track on Datasets and Benchmarks}, volume~1, 2021.
\newblock URL \url{https://datasets-benchmarks-proceedings.neurips.cc/paper_files/paper/2021/file/158f3069a435b314a80bdcb024f8e422-Paper-round2.pdf}.

\bibitem[Luecken et~al.(2022)Luecken, B{\"u}ttner, Chaichoompu, Danese, Interlandi, M{\"u}ller, Strobl, Zappia, Dugas, Colom{\'e}-Tatch{\'e}, et~al.]{luecken2022benchmarking}
Luecken, M.~D., B{\"u}ttner, M., Chaichoompu, K., Danese, A., Interlandi, M., M{\"u}ller, M.~F., Strobl, D.~C., Zappia, L., Dugas, M., Colom{\'e}-Tatch{\'e}, M., et~al.
\newblock {Benchmarking atlas-level data integration in single-cell genomics}.
\newblock \emph{Nature Methods}, 19\penalty0 (1):\penalty0 41--50, 2022.

\bibitem[Marks et~al.(2025)Marks, Knott, Kondapaneni, Cole, Defraeye, Perez-Cruz, and Perona]{marks2024closerlookbenchmarkingselfsupervised}
Marks, M., Knott, M., Kondapaneni, N., Cole, E., Defraeye, T., Perez-Cruz, F., and Perona, P.
\newblock A closer look at benchmarking self-supervised pre-training with image classification.
\newblock \emph{International Journal of Computer Vision}, Apr 2025.
\newblock ISSN 1573-1405.
\newblock \doi{10.1007/s11263-025-02402-w}.
\newblock URL \url{https://doi.org/10.1007/s11263-025-02402-w}.

\bibitem[Min et~al.(2023)Min, Ross, Sulem, Veyseh, Nguyen, Sainz, Agirre, Heintz, and Roth]{min2023recent}
Min, B., Ross, H., Sulem, E., Veyseh, A. P.~B., Nguyen, T.~H., Sainz, O., Agirre, E., Heintz, I., and Roth, D.
\newblock {Recent advances in natural language processing via large pre-trained language models: A survey}.
\newblock \emph{ACM Computing Surveys}, 56\penalty0 (2):\penalty0 1--40, 2023.
\newblock \doi{10.1145/3605943}.
\newblock URL \url{https://doi.org/10.1145/3605943}.

\bibitem[Pearson(1901)]{pearson1901liii}
Pearson, K.
\newblock {LIII. On lines and planes of closest fit to systems of points in space}.
\newblock \emph{The London, Edinburgh, and Dublin philosophical magazine and journal of science}, 2\penalty0 (11):\penalty0 559--572, 1901.

\bibitem[Pola{\'n}ski et~al.(2019)Pola{\'n}ski, Young, Miao, Meyer, Teichmann, and Park]{polanski2019bbknn}
Pola{\'n}ski, K., Young, M.~D., Miao, Z., Meyer, K.~B., Teichmann, S.~A., and Park, J.-E.
\newblock {BBKNN: Fast Batch Alignment of Single Cell Transcriptomes}.
\newblock \emph{Bioinformatics}, 2019.
\newblock \doi{10.1093/bioinformatics/btz625}.

\bibitem[Radford et~al.(2021)Radford, Kim, Hallacy, Ramesh, Goh, Agarwal, Sastry, Askell, Mishkin, Clark, et~al.]{CLIP}
Radford, A., Kim, J.~W., Hallacy, C., Ramesh, A., Goh, G., Agarwal, S., Sastry, G., Askell, A., Mishkin, P., Clark, J., et~al.
\newblock {Learning transferable visual models from natural language supervision}.
\newblock In \emph{ICML}, pp.\  8748--8763, 2021.

\bibitem[Richter et~al.(2024{\natexlab{a}})Richter, Bahrami, Xia, Fischer, and Theis]{Richter2024}
Richter, T., Bahrami, M., Xia, Y., Fischer, D.~S., and Theis, F.~J.
\newblock {Delineating the effective use of self-supervised learning in single-cell genomics}.
\newblock \emph{Nature Machine Intelligence}, Dec 2024{\natexlab{a}}.
\newblock ISSN 2522-5839.
\newblock \doi{10.1038/s42256-024-00934-3}.
\newblock URL \url{https://doi.org/10.1038/s42256-024-00934-3}.

\bibitem[Richter et~al.(2024{\natexlab{b}})Richter, Bahrami, Xia, Fischer, and Theis]{Richter2024.02.16.580624}
Richter, T., Bahrami, M., Xia, Y., Fischer, D.~S., and Theis, F.~J.
\newblock {Delineating the Effective Use of Self-Supervised Learning in Single-Cell Genomics}.
\newblock \emph{bioRxiv}, 2024{\natexlab{b}}.
\newblock \doi{10.1101/2024.02.16.580624}.
\newblock URL \url{https://www.biorxiv.org/content/early/2024/02/18/2024.02.16.580624}.

\bibitem[Schiappa et~al.(2023)Schiappa, Rawat, and Shah]{schiappa2023self}
Schiappa, M.~C., Rawat, Y.~S., and Shah, M.
\newblock {Self-supervised learning for videos: A survey}.
\newblock \emph{ACM Computing Surveys}, 55\penalty0 (13s):\penalty0 1--37, 2023.
\newblock \doi{10.1145/3577925}.
\newblock URL \url{https://doi.org/10.1145/3577925}.

\bibitem[Sikkema et~al.(2023)Sikkema, Ram{\'i}rez-Su{\'a}stegui, Strobl, Gillett, Zappia, Madissoon, et~al.]{sikkema2022integrated}
Sikkema, L., Ram{\'i}rez-Su{\'a}stegui, C., Strobl, D.~C., Gillett, T.~E., Zappia, L., Madissoon, E., et~al.
\newblock {An integrated cell atlas of the lung in health and disease}.
\newblock \emph{{Nature Medicine}}, 29\penalty0 (6):\penalty0 1563--1577, Jun 2023.
\newblock ISSN 1546-170X.
\newblock \doi{10.1038/s41591-023-02327-2}.
\newblock URL \url{https://doi.org/10.1038/s41591-023-02327-2}.

\bibitem[Slyper et~al.(2020)Slyper, Porter, Ashenberg, Waldman, Drokhlyansky, Wakiro, Smillie, Smith-Rosario, Wu, Dionne, et~al.]{slyper2020single}
Slyper, M., Porter, C.~B., Ashenberg, O., Waldman, J., Drokhlyansky, E., Wakiro, I., Smillie, C., Smith-Rosario, G., Wu, J., Dionne, D., et~al.
\newblock {A single-cell and single-nucleus RNA-Seq toolbox for fresh and frozen human tumors}.
\newblock \emph{{Nature Medicine}}, 26\penalty0 (5):\penalty0 792--802, 2020.

\bibitem[Sohn(2016)]{ntxent}
Sohn, K.
\newblock {Improved deep metric learning with multi-class N-pair loss objective}.
\newblock In \emph{NeurIPS}, pp.\  1857–1865, 2016.
\newblock ISBN 9781510838819.

\bibitem[Stoeckius et~al.(2017)Stoeckius, Hafemeister, Stephenson, Houck-Loomis, Chattopadhyay, Swerdlow, Satija, and Smibert]{CITE-seq}
Stoeckius, M., Hafemeister, C., Stephenson, W., Houck-Loomis, B., Chattopadhyay, P.~K., Swerdlow, H., Satija, R., and Smibert, P.
\newblock {Simultaneous epitope and transcriptome measurement in single cells}.
\newblock \emph{Nature Methods}, 14\penalty0 (9):\penalty0 865--868, Sep 2017.
\newblock ISSN 1548-7105.
\newblock \doi{10.1038/nmeth.4380}.
\newblock URL \url{https://doi.org/10.1038/nmeth.4380}.

\bibitem[Susmelj et~al.(2023)Susmelj, Heller, Wirth, Prescott, Ebner, and {et al.}]{Susmelj_Lightly}
Susmelj, I., Heller, M., Wirth, P., Prescott, J., Ebner, M., and {et al.}
\newblock {Lightly}, 2023.
\newblock URL \url{https://github.com/lightly-ai/lightly}.

\bibitem[Swanson et~al.(2020)Swanson, Lord, Reading, Heubeck, Savage, Green, Li, Torgerson, Bumol, Graybuck, and Skene]{TEA-seq}
Swanson, E., Lord, C., Reading, J., Heubeck, A.~T., Savage, A.~K., Green, R., Li, X.-j., Torgerson, T.~R., Bumol, T.~F., Graybuck, L.~T., and Skene, P.~J.
\newblock {TEA-seq: a trimodal assay for integrated single cell measurement of transcription, epitopes, and chromatin accessibility}.
\newblock \emph{bioRxiv}, 2020.
\newblock \doi{10.1101/2020.09.04.283887}.
\newblock URL \url{https://www.biorxiv.org/content/early/2020/11/16/2020.09.04.283887}.

\bibitem[Tang et~al.(2009)Tang, Barbacioru, Wang, Nordman, Lee, Xu, Wang, Bodeau, Tuch, Siddiqui, Lao, and Surani]{scRNA-seq}
Tang, F., Barbacioru, C., Wang, Y., Nordman, E., Lee, C., Xu, N., Wang, X., Bodeau, J., Tuch, B.~B., Siddiqui, A., Lao, K., and Surani, M.~A.
\newblock {mRNA-Seq whole-transcriptome analysis of a single cell}.
\newblock \emph{Nature Methods}, 6\penalty0 (5):\penalty0 377--382, May 2009.
\newblock ISSN 1548-7105.
\newblock \doi{10.1038/nmeth.1315}.
\newblock URL \url{https://doi.org/10.1038/nmeth.1315}.

\bibitem[Tang et~al.(2024)Tang, Chen, Chen, Yao, You, and Chen]{monae}
Tang, Z., Chen, G., Chen, S., Yao, J., You, L., and Chen, C. Y.-C.
\newblock {Modal-nexus auto-encoder for multi-modality cellular data integration and imputation}.
\newblock \emph{{Nature Communications}}, 15\penalty0 (1):\penalty0 9021, Oct 2024.
\newblock ISSN 2041-1723.
\newblock \doi{10.1038/s41467-024-53355-6}.
\newblock URL \url{https://doi.org/10.1038/s41467-024-53355-6}.

\bibitem[Theodoris et~al.(2023)Theodoris, Xiao, Chopra, Chaffin, Al~Sayed, Hill, Mantineo, Brydon, Zeng, Liu, et~al.]{theodoris2023transfer}
Theodoris, C.~V., Xiao, L., Chopra, A., Chaffin, M.~D., Al~Sayed, Z.~R., Hill, M.~C., Mantineo, H., Brydon, E.~M., Zeng, Z., Liu, X.~S., et~al.
\newblock {Transfer learning enables predictions in network biology}.
\newblock \emph{Nature}, 618\penalty0 (7965):\penalty0 616--624, 2023.

\bibitem[Toma et~al.(2024)Toma, Ovcharenko, Daunhawer, Vogt, Barkmann, and Boeva]{toma2024benchmarking}
Toma, P., Ovcharenko, O., Daunhawer, I., Vogt, J.~E., Barkmann, F., and Boeva, V.
\newblock Benchmarking self-supervised learning for single-cell data.
\newblock In \emph{NeurIPS 2024 Workshop: Self-Supervised Learning-Theory and Practice}, 2024.

\bibitem[Tran et~al.(2020)Tran, Ang, Chevrier, Zhang, Lee, Goh, and Chen]{Tran2020}
Tran, H. T.~N., Ang, K.~S., Chevrier, M., Zhang, X., Lee, N. Y.~S., Goh, M., and Chen, J.
\newblock {A benchmark of batch-effect correction methods for single-cell RNA sequencing data}.
\newblock \emph{Genome Biology}, 21\penalty0 (1):\penalty0 12, Jan 2020.
\newblock ISSN 1474-760X.
\newblock \doi{10.1186/s13059-019-1850-9}.
\newblock URL \url{https://doi.org/10.1186/s13059-019-1850-9}.

\bibitem[van~den Oord et~al.(2019)van~den Oord, Li, and Vinyals]{infonce}
van~den Oord, A., Li, Y., and Vinyals, O.
\newblock {Representation Learning with Contrastive Predictive Coding}, 2019.
\newblock URL \url{https://arxiv.org/abs/1807.03748}.

\bibitem[Vaswani et~al.(2017)Vaswani, Shazeer, Parmar, Uszkoreit, Jones, Gomez, Kaiser, and Polosukhin]{vaswani2017attention}
Vaswani, A., Shazeer, N., Parmar, N., Uszkoreit, J., Jones, L., Gomez, A.~N., Kaiser, {\L}., and Polosukhin, I.
\newblock {Attention is all you need}.
\newblock \emph{Advances in neural information processing systems}, 30, 2017.

\bibitem[Wang et~al.(2019)Wang, Ma, Chen, Luo, Yi, and Bailey]{wang2019symmetriccrossentropyrobust}
Wang, Y., Ma, X., Chen, Z., Luo, Y., Yi, J., and Bailey, J.
\newblock Symmetric cross entropy for robust learning with noisy labels.
\newblock In \emph{Proceedings of the IEEE/CVF international conference on computer vision}, pp.\  322--330, 2019.

\bibitem[Wolf et~al.(2018)Wolf, Angerer, and Theis]{Wolf2018-mv}
Wolf, F.~A., Angerer, P., and Theis, F.~J.
\newblock {SCANPY: large-scale single-cell gene expression data analysis}.
\newblock \emph{Genome Biology}, 19\penalty0 (1):\penalty0 15, Feb 2018.
\newblock ISSN 1474-760X.
\newblock \doi{10.1186/s13059-017-1382-0}.
\newblock URL \url{https://doi.org/10.1186/s13059-017-1382-0}.

\bibitem[Xiong et~al.(2023)Xiong, Chen, and Kellis]{scCLIP}
Xiong, L., Chen, T., and Kellis, M.
\newblock {scCLIP: Multi-modal Single-cell Contrastive Learning Integration Pre-training}.
\newblock In \emph{NeurIPS 2023 AI for Science Workshop}, 2023.
\newblock URL \url{https://openreview.net/forum?id=KMtM5ZHxct}.

\bibitem[Xu et~al.(2021)Xu, Lopez, Mehlman, Regier, Jordan, and Yosef]{https://doi.org/10.15252/msb.20209620}
Xu, C., Lopez, R., Mehlman, E., Regier, J., Jordan, M.~I., and Yosef, N.
\newblock {Probabilistic harmonization and annotation of single‐cell transcriptomics data with deep generative models}.
\newblock \emph{Molecular Systems Biology}, 17\penalty0 (1):\penalty0 e9620, 2021.
\newblock \doi{https://doi.org/10.15252/msb.20209620}.
\newblock URL \url{https://www.embopress.org/doi/abs/10.15252/msb.20209620}.

\bibitem[Xue et~al.(2024)Xue, Gan, Ni, Joshi, and Mirzasoleiman]{xue2024investigatingbenefitsprojectionhead}
Xue, Y., Gan, E., Ni, J., Joshi, S., and Mirzasoleiman, B.
\newblock {Investigating the Benefits of Projection Head for Representation Learning}.
\newblock In \emph{ICLR}, 2024.
\newblock URL \url{https://arxiv.org/abs/2403.11391}.

\bibitem[Yan et~al.(2023)Yan, Zheng, Wu, and Li]{XuhuaRFL23}
Yan, X., Zheng, R., Wu, F., and Li, M.
\newblock {CLAIRE: contrastive learning-based batch correction framework for better balance between batch mixing and preservation of cellular heterogeneity}.
\newblock \emph{Bioinformatics (Oxford, England)}, 39, 02 2023.
\newblock \doi{10.1093/bioinformatics/btad099}.

\bibitem[Yang et~al.(2022{\natexlab{a}})Yang, Wang, Wang, Fang, Tang, Huang, Lu, and Yao]{yang2022scbert}
Yang, F., Wang, W., Wang, F., Fang, Y., Tang, D., Huang, J., Lu, H., and Yao, J.
\newblock {scBERT as a large-scale pretrained deep language model for cell type annotation of single-cell RNA-seq data}.
\newblock \emph{Nature Machine Intelligence}, 4\penalty0 (10):\penalty0 852--866, 2022{\natexlab{a}}.

\bibitem[Yang et~al.(2022{\natexlab{b}})Yang, Yang, Xie, Ni, Liu, Yang, Mu, and Wang]{Concerto}
Yang, M., Yang, Y., Xie, C., Ni, M., Liu, J., Yang, H., Mu, F., and Wang, J.
\newblock {Contrastive learning enables rapid mapping to multimodal single-cell atlas of multimillion scale}.
\newblock \emph{Nature Machine Intelligence}, 4\penalty0 (8):\penalty0 696--709, 2022{\natexlab{b}}.

\bibitem[Yu et~al.(2023)Yu, Xu, Zhang, and Li]{Yu2023}
Yu, X., Xu, X., Zhang, J., and Li, X.
\newblock {Batch alignment of single-cell transcriptomics data using deep metric learning}.
\newblock \emph{{Nature Communications}}, 14\penalty0 (1):\penalty0 960, 2023.
\newblock \doi{10.1038/s41467-023-36635-5}.

\bibitem[Zbontar et~al.(2021)Zbontar, Jing, Misra, LeCun, and Deny]{zbontar2021barlowtwinsselfsupervisedlearning}
Zbontar, J., Jing, L., Misra, I., LeCun, Y., and Deny, S.
\newblock Barlow twins: Self-supervised learning via redundancy reduction.
\newblock In \emph{International conference on machine learning}, pp.\  12310--12320. PMLR, 2021.

\bibitem[Zhang \& Ma(2022)Zhang and Ma]{zhang2022rethinkingaugmentationmodulecontrastive}
Zhang, J. and Ma, K.
\newblock Rethinking the augmentation module in contrastive learning: Learning hierarchical augmentation invariance with expanded views.
\newblock In \emph{Proceedings of the IEEE/CVF Conference on Computer Vision and Pattern Recognition}, pp.\  16650--16659, 2022.

\bibitem[Zhang et~al.(2024)Zhang, Mathew, Lim, Mason, Martinez, Huang, Wherry, Susztak, Minn, Ma, and Zhang]{Zhang2023DT}
Zhang, Z., Mathew, D., Lim, T.~L., Mason, K., Martinez, C.~M., Huang, S., Wherry, E.~J., Susztak, K., Minn, A.~J., Ma, Z., and Zhang, N.~R.
\newblock Recovery of biological signals lost in single-cell batch integration with cellanova.
\newblock \emph{{Nature Biotechnology}}, Nov 2024.
\newblock ISSN 1546-1696.
\newblock \doi{10.1038/s41587-024-02463-1}.
\newblock URL \url{https://doi.org/10.1038/s41587-024-02463-1}.

\end{thebibliography}
\bibliographystyle{icml2025}

\newpage
\appendix
\onecolumn
\section{Generic Self-Supervised Methods}
\label{sec:generic_methods}

In contrastive SSL, different augmentations/modalities of the same instance are used to create positive pairs (i.e., similar examples), while pairs of distinct instances represent negative pairs (i.e., dissimilar examples)~\cite{chen2020simple}. 
Non-contrastive methods, also called negative-free contrastive learning, leverage only positive pairs~\cite{CaoNegativeFree}.

\paragraph{Contrastive Methods:} A common framework for contrastive learning is SimCLR~\cite{chen2020simple}. Originally, SimCLR applies three image-data-specific augmentations to create positive/negative pairs and maximizes agreement between different augmented views via a temperature-scaled cross-entropy loss (NTXent)~\cite{ntxent} loss.
MoCo~\cite{he2020momentum}
uses an additional momentum encoder for the augmented views. The key advantage of Moco is improved memory efficiency.
NNCLR~\cite{dwibedi2021littlehelpfriendsnearestneighbor} samples nearest neighbors of an instance to define positive pairs.
Both MoCo and NNCLR contrast via InfoNCE~\cite{infonce} loss.
SimCLR~\cite{chen2020simple}, MoCo~\cite{he2020momentum}, and NNCLR~\cite{dwibedi2021littlehelpfriendsnearestneighbor} rely on positive and negative samples.

\paragraph{Non-Contrastive Methods:} 
The emergence of non-contrastive methods was facilitated by an improved understanding of instabilities during model training.
BYOL~\cite{grill2020bootstrap} uses online (trainable) and target (fixed) networks to train a representation. 
Online and target networks receive actual samples and augmented views respectively. 
The target network is updated using exponential moving averages of weights of the previous online networks.
BYOL~\cite{grill2020bootstrap} minimizes the similarity of two representations produced by both networks.
Additionally, using momentum encoders in MoCo~\cite{he2020momentum} and BYOL~\cite{grill2020bootstrap} helps against dimensionality collapse from, for instance, the lack of negative pairs.
Dimensionality collapse appears when embeddings span in a lower-dimensional subspace instead of the entire space~\cite{jing2022understanding}.
SimSiam, like SimCLR, uses an encoder with shared weights to process two augmented views.
One encoder output (\textit{left}) is fed to an additional predictor network before maximizing similarity, while the other output (\textit{right}) is used directly. 
The encoder and predictor are updated with only \textit{left} path gradients.
SimSiam~\cite{chen2020exploringsimplesiameserepresentation} and BYOL~\cite{grill2020bootstrap} use a predictor network to achieve better performance and avoid representation collapse without leveraging negative pairs.
Barlow~Twins~\cite{zbontar2021barlowtwinsselfsupervisedlearning} uses augmentations to train two identical networks and calculates a cross-correlation matrix~\cite{zbontar2021barlowtwinsselfsupervisedlearning} between the trained representations to reduce redundancy of the embeddings. 
VICReg~\cite{vicreg} is a joint embedding architecture with variance, invariance, and covariance regularization. 
The authors introduce regularization terms to the loss to control the variance of embeddings and decorrelate latent variables.





\section{Datasets}
\label{datasets}



All datasets used in our benchmark are publicly available.

\textbf{Human Immune Cells (HIC):} This dataset comprises 33,506 cells and includes 12,303 genes from ten different donors assembled by~\citeauthor{luecken2022benchmarking}(\citeyear{luecken2022benchmarking}) from five studies. One study derived cells from the human bone marrow and the other four from the human peripheral blood. There are 16 cell types annotated in the dataset. 
Availability: \url{https://doi.org/10.6084/m9.figshare.12420968.v8}

\textbf{Mouse Cell Atlas (MCA):} This dataset comprises 6,954 cells collected across two studies~\cite{Tran2020} 
with the first study consisting of 4,239 cells and the second batch containing 2,715 cells. Three different sequencing protocols were used. The harmonized dataset contains 51,817 genes and eleven cell types. 
Availability: \url{https://ndownloader.figshare.com/files/10351110} and \url{https://ndownloader.figshare.com/files/10760158}

\textbf{Peripheral Blood Mononuclear Cells (PBMC):} Collected by \citeauthor{Ding632216}(\citeyear{Ding632216}), this dataset contains 30,449 cells from two patients and includes 33,694 genes. Cells were sequenced with seven different protocols (10x Chromium (v2), 10x Chromium (v3), Drop-seq, inDrops, Chromium (v3), Seq-Well, CEL-Seq2). We have made use of the annotations of nine unique cell types (D4+ T cell, Cytotoxic T cell, Natural killer cell, CD16+ monocyte, CD14+ monocyte, Megakaryocyte, B cell, Dendritic cell, Plasmacytoid dendritic cell) provided in the original study. Also, we removed the unassigned cells. 
Availability: \url{https://singlecell.broadinstitute.org/single_cell/study/SCP424/single-cell-comparison-pbmc-data}

\textbf{Pancreas:} This dataset was collected by \citeauthor{Tran2020}(\citeyear{Tran2020}) combining five studies of the human pancreas. It comprises 14,767 cells, with 5,975 genes shared across all studies, sequenced by four scRNA-seq technologies (inDrop, CEL-Seq2, Smart-Seq2, SMARTer). The harmonized dataset contains 13 cell types (alpha, beta, ductal, acinar, delta, pancreatic stellate, pancreatic polypeptide, endothelial, macrophage, mast, epsilon, Schwann and T cell).
Availability: \url{https://figshare.com/ndownloader/files/24539828}

\textbf{Lung:} This dataset contains 32,426 cells across 16 batches and two technologies (Drop-seq and 10x Chromium), assembled by~\citeauthor{luecken2022benchmarking}(\citeyear{luecken2022benchmarking}) from three labs. The harmonized dataset includes 15,148 genes. The cells are derived from transplant patients and lung biopsies and are annotated as 17 cell types.
Availability: \url{https://figshare.com/ndownloader/files/24539942}

\textbf{Immune Cell Atlas:} This dataset contains 329,762 cells and includes 36,398 genes across twelve batches and three different sequencing technologies (10x 5' v1, 10x 5' v2, 10x 3' v3), collected by~\citeauthor{doi:10.1126/science.abl5197}(\citeyear{doi:10.1126/science.abl5197}). The cells originate from 16 different tissues. The annotations include 35 fine-grain cell types.
Availability: \url{https://datasets.cellxgene.cziscience.com/08f58b32-a01b-4300-8ebc-2b93c18f26f7.h5ad}

\textbf{Tabula Sapiens:} This dataset was collected by~\citeauthor{tabula}(\citeyear{tabula}) and contains 1,136,218 cells from 24 tissues and organs, sequenced with 10x 3' v3, 10x 5' v2, Smart-seq, and Smart-seq3 protocols. 
Tabula Sapiens is a molecular reference atlas for more than 400 cell types of the human body.
Availability: \url{https://cellxgene.cziscience.com/collections/e5f58829-1a66-40b5-a624-9046778e74f5}

\textbf{Multi-modal Peripheral Blood Mononuclear Cells (PBMC-M):} This dataset was collected by \citeauthor{HAO20213573}(\citeyear{HAO20213573}) with 161,764 cells across eight batches. For each cell, two modalities are available: RNA and protein. RNA has 18,702 genes, while the dimension of protein is 224.
As a pre-processing step, we merge different T cell granularities, similar to the Concerto framework~\cite{Concerto}.
Availability: \url{https://atlas.fredhutch.org/data/nygc/multimodal/pbmc_multimodal.h5seurat}

\textbf{Multi-modal Bone Marrow Mononuclear Cells (BMMC).} This dataset was collected by \citeauthor{NeuripsDataChallenge}(\citeyear{NeuripsDataChallenge}) and contains 90,261 cells across thirteen batches and twelwe healthy human donors~\cite{NeuripsDataChallengePaper}. 
Each cell has two modalities: Gene expression (GEX) and protein abundance (ADT). While GEX has 13,953 genes, the protein abundance dimension is 134.
Pre-processing is the same as PBMC-M. 
Availability: \url{https://www.ncbi.nlm.nih.gov/geo/query/acc.cgi?acc=GSE194122}

\section{Evaluation Details}
\label{sec:eval_details}

\paragraph{Preprocessing:} All datasets are preprocessed using \textsc{scanpy}~\cite{Wolf2018-mv} normalize-total function, which scales the total counts per cell to 10,000, followed by log-transformation. We subsequently perform batch-aware feature selection to choose the 4,000 most highly-variable genes (HVGs) for further processing.
For multi-modal PBMC-M and BMMC datasets, we select 2,000 HVGs contrary to 4,000 HVGs for the single modality datasets.

\paragraph{Batch Correction:} 
The evaluated metrics are divided into two categories: those that measure the conservation of biological variance and that that measure the batch correction~\cite{Tran2020, luecken2022benchmarking}.
To evaluate conservation of biological variation, we calculate the isolated labels score, the Leiden NMI and ARI, the silhouette label score, and the cLISI metric.
To evaluate batch correction, we calculate the graph connectivity, kBET per label, iLISI for each cell,  the PCR comparison score, and the silhouette coefficient per batch.
For details and definitions of the used evaluation metrics, as well as their implementation, we refer to \cite{luecken2022benchmarking}.

\paragraph{Cell-Type Annotation and Missing Modality Prediction:}
In the PBMC-M dataset, for cell-type annotation mapping and missing modality inference, we hold out batches \textit{P3}, \textit{P5}, and \textit{P8}.
In the BMMC dataset, for cell-type annotation mapping and missing modality inference, we hold out batches \textit{s4d1}, \textit{s4d8}, and \textit{s4d9}.
Similar to the approach of~\cite{https://doi.org/10.15252/msb.20209620}, we perform cell-type annotation by fitting a non-parametric supervised classifier (k-nearest neighbors (KNN) classifier with $k=11$). For missing modality prediction, we fit a KNN classifier with $k=5$, as in~\cite{Concerto}.

\section{Hyperparameter Tuning}
\label{sec:hparams}
In all experiments, we use the augmentation pipeline proposed by CLEAR~\cite{CLEAR} as a foundation, unless stated differently. Experiments described in this section were computed for all methods except Concerto. For the latter, we use the original model from~\cite{Concerto}.

\paragraph{Optimization:} All models in this benchmark, except Concerto, were trained with the Adam optimizer~\cite{kingma2017adammethodstochasticoptimization}.
We use a stepwise learning rate schedule with base learning rate \verb|1e-4| and fix the batch size at 256. 
When applicable, the memory bank size was set to 2048.

\paragraph{Encoder Architecture:}\label{encoder-def} 
We fix the encoder across all architectures and only perform a hyper-parameter search on the dimensionality of the encoder output, i.e., the representation dimensionality. The encoder consists of a fully connected layer reducing the dimensionality to 128, followed by a ReLU activation and batch normalization. A further fully connected layer encodes the hidden representation to the representation dimension, followed by batch normalization.


\paragraph{Projector Dimensionality:}\label{projection-def}
Projection heads benefit self-supervised models in learning robust representations~\cite{xue2024investigatingbenefitsprojectionhead}. At inference, the projection head is discarded, and only the (backbone) encoder is used for inference. All evaluated architectures subject to our evaluation include a projection head. We perform a hyperparameter search to find the best output dimension of the projector. 

All projection heads were implemented as noted in the respective works. In their respective works, SimCLR, MoCo, SimSiam, and NNCLR are evaluated with projectors that retain or scale down the dimensionality of the representation. BarlowTwins, BYOL, and VICReg are evaluated with projectors that retain or scale up the dimensionality. We follow this rationale and search a grid of scaling factors $\{1, 2, 4\}$. To compute the projection dimensionality, the scaling factor is either divided (scale-down models) or multiplied (scale-up models) with the representation's dimension.

\paragraph{Regularization Hyperparameters:} Variance-invariance-covariance regularization hyperparameters are used as is done in the original work. We evaluate a grid of parameters, where the invariance term and the variance term $\lambda, \alpha = \{5, 10, 25, 50\}$, while the invariance term $\beta$ is fixed to 1. We find that $\lambda$ and $\alpha$ fixed to 5 perform well across both ablation datasets.

\paragraph{Augmentation Strength:} Augmentations are known to benefit SSL models in finding robust representations. Details of the evaluated augmentations are listed in~\autoref{sec:aug-ablation}. We perform a grid search to optimize the hyperparameters for all augmentations.
This includes $\alpha$ for all models, $\sigma$ for the Gaussian Noise augmentation, and the \textit{KNN}-size for the nearest-neighbor-based transforms MNN and BBKNN. 
For each augmentation, the original CLEAR hyperparameters are fixed, and only the hyperparameters of the evaluated augmentation are adapted.
For the ablation of BBKNN, we remove CrossOver, and replace it by BBKNN. Due to the implementation of MNN, we remove CrossOver and insert MNN at the front of the augmentation pipeline.
Results of the ablation are recorded in~\autoref{fig:alpha-ablation}.

\section{Augmentations}
\label{sec:aug-ablation}


We evaluate six augmentations in this work. For all, the parameter $\alpha$ defines the proportion of values affected by the transform. Augmentations are applied sequentially.
Masking is performed by setting gene expressions to zero. Gaussian noise computes a noise vector computed from the normal distribution (with zero mean and standard deviation $\sigma$) and adds it to the input. InnerSwap switches expressions between genes within a cell, while CrossOver switches expressions of the same gene between two \textit{random} cells.

The MNN augmentation refers to our implementation of CLAIRE's augmentation~\cite{XuhuaRFL23}. For each cell, it computes an intra- and inter-batch neighborhood based on its mutual nearest neighbors. Then, views are computed by interpolating between neighbors. We do not filter cell-neighborhoods based on representation similarities during early stages of model training, as is done in the original work.
This work introduces the BBKNN augmentation. It uses a non-trimmed batch-balanced KNN graph~\cite{polanski2019bbknn} to define a set of \textit{\#batches}$\times$\textit{KNN} neighbors for each cell. Views are computed by interpolating between neighbors. It differs from CLAIRE's concept in that it does not distinguish between intra- and inter-batch neighbors. While MNN always produces a view based on neighbors within and a view based on neighbors from outside the batch, this is not the case for BBKNN.
Due to its implementation, the MNN augmentation is limited to be applied first in any augmentation pipeline. We refer to~\cite{XuhuaRFL23} for further detail on the interpolation process.

\section{Multi-Modal Setting}
\label{sec:multimodal_setting}

Recent developments in the single-cell analysis allow the measurement of multiple aspects of a cellular state.
Data containing multiple modalities of a cell, e.g., RNA and protein, is called multi-omics.
Existing self-supervised methods for single-cell data integration can be extended to the multimodal setting by combining views produced by specialized models for different modalities.
We train two models for each modality; each model consists of an encoder and projector.
As is common~\cite{he2020momentum, chen2020simple, geiping2023cookbook}, only the encoder is used to infer the integrated representation.
However, in the single-cell community, the projector is also used during inference, and, therefore, we also evaluate whether projection during prediction improves performance.
Additionally, there are various techniques to combine representations~\cite{CLIP, scCLIP, Concerto}. We evaluate three approaches: Addition, concatenation, and CLIP.


\paragraph{Encoder \& Projector Embedding Evaluation:}
Using CLEAR augmentations, we train two models for each modality, each consisting of an encoder and projector.
In~\autoref{tab:multimodal_projection_bc}, we compare data integration performance with and without a projector during inference.
Interestingly, SimCLR benefits from projection, while VICReg performance degrades.
We conclude that the effect of projection is inconsistent across models.


\renewcommand{\thefigure}{G\arabic{figure}}
\setcounter{figure}{0}

\section{Supplementary Figures}

\begin{figure}[ht!]
    \centering
    \includegraphics[scale=1.01]{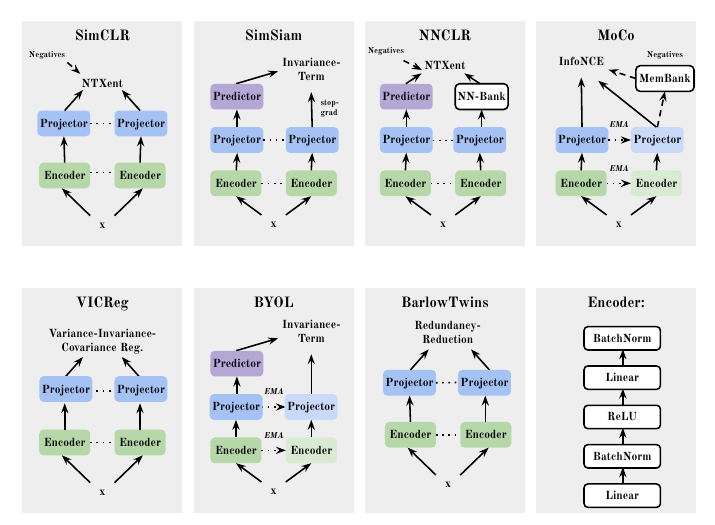}
    \caption{Overview of considered methods. Dotted lines between the encoder and projector blocks represent weight sharing. Exponential Moving Average (EMA) denotes the updating of weights with momentum. This figure was inspired by~\protect\cite{vicreg, he2020momentum, dwibedi2021littlehelpfriendsnearestneighbor} based on our implementation of models with LightlySSL~\cite{Susmelj_Lightly}.}
    \label{fig:Model-Overview}
\end{figure}

\begin{figure}[ht!]
    \centering
    \minipage{0.48\textwidth}
        \input{figures/downstream/batch_correct}
    \endminipage
    \hfill
    \minipage{0.48\textwidth}
        \input{figures/downstream/query_to_reference}
    \endminipage
\end{figure}

\begin{figure}[t!]
\centering
\includegraphics[width=0.8\textwidth]{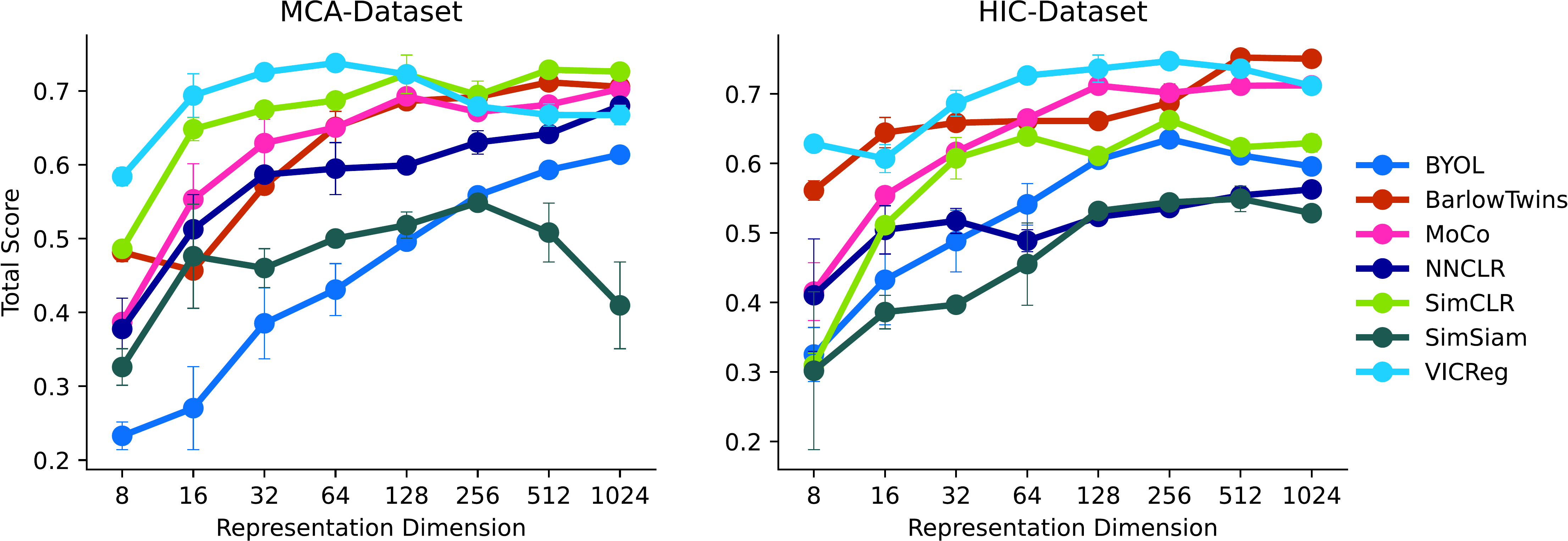}
\caption{Tuning of the encoder based on the representation 
dimensionality. The encoder architecture is defined in~\autoref{encoder-def}. Lines correspond to the mean total score across five runs with unique seeds.}
\label{fig:rep-dim}
\end{figure}

\begin{figure}[t!]
\centering
\includegraphics[width=0.8\textwidth]{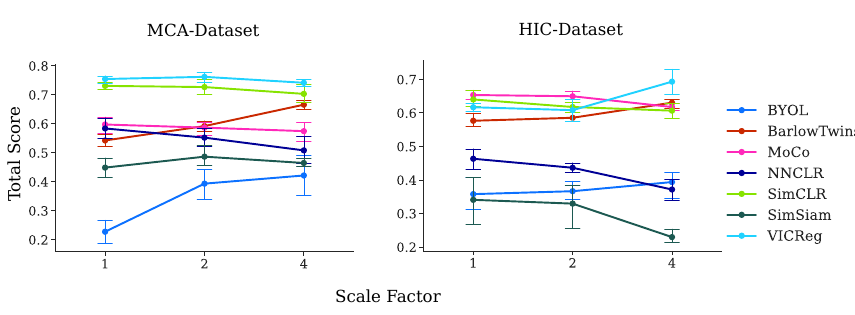}
\caption{Tuning of the projector. The scale factor is defined in~\autoref{projection-def}: in contrastive methods, the projected size decreased according to the scale factor, while for non-contrastive methods the projection size increases in accordance with the scale factor. Lines correspond to the mean total score across five runs with unique seeds.}
\label{fig:projection-dim}
\end{figure}

\begin{figure}[H]
\centering
\includegraphics[width=0.8\linewidth]{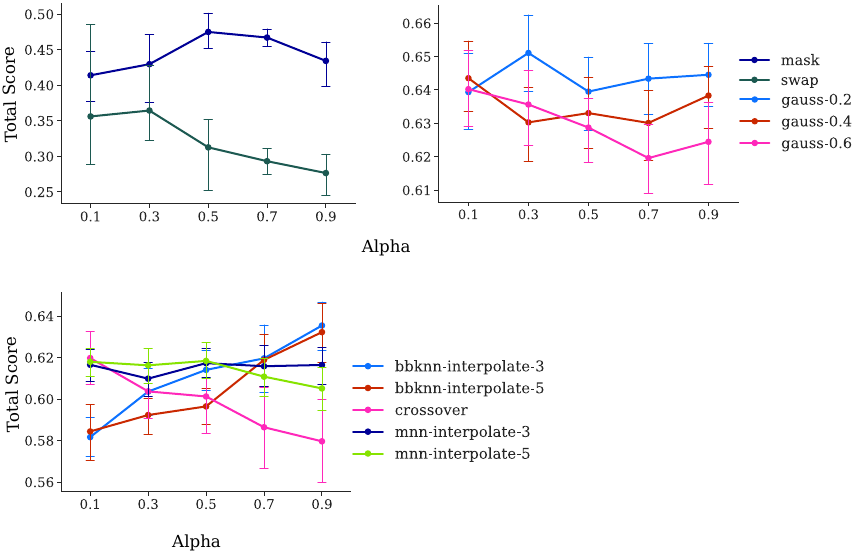}
\caption{Ablation on the augmentation hyperparameters. The figure aggregates results for all methods, trained on the HIC dataset.}
\label{fig:alpha-ablation}
\end{figure}


  

      


\begin{figure}[H]
    \centering
    \begin{subfigure}[t!]{0.5\linewidth}
        \centering
        \includegraphics[]{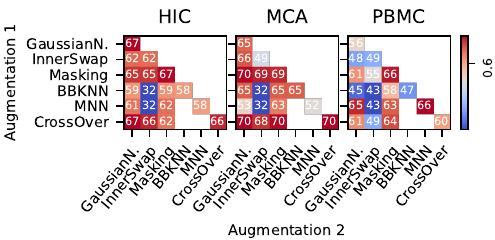}
        \caption{SimCLR}
    \end{subfigure}%
    ~ 
    \begin{subfigure}[t!]{0.5\linewidth}
        \centering
        \includegraphics[]{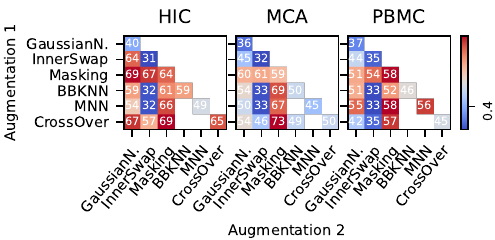}
        \caption{MoCo}
    \end{subfigure}
    \vspace{-0.3cm}
    
    \caption{Evaluation of individual and combined data augmentations based on total score for batch correction for SimCLR and MoCo method. Diagonal entries correspond to a single augmentation, and off-diagonal entries correspond to the two sequentially applied augmentations.}
  \label{fig:heatmaps_aug_more}
\end{figure}


\section{Supplementary Tables}
\label{sec:supplementary-tables}

\setcounter{table}{0}
\renewcommand{\thetable}{H\arabic{table}}

\begin{table}[H]
\centering
\caption{Batch correction benchmark for methods trained using the CLEAR pipeline. This table is an extension to~\autoref{tab:clear_bc}, containing datasets that were used during hyperparameter tuning.}
\vspace{0.3cm}
\setlength\tabcolsep{10pt}
\begin{tabular}{l|ccc|ccc|ccc}
\toprule
\multirow{2}{*}{\textbf{Method}} & \multicolumn{3}{c|}{\textbf{MCA}} & \multicolumn{3}{c|}{\textbf{HIC}} & \multicolumn{3}{c}{\textbf{Lung}} \\
& Bio & Batch & Total & Bio & Batch & Total & Bio & Batch & Total \\
\midrule
\multirow{2}{*}{SimCLR} & 0.519 & 0.666 & 0.578 & 0.753 & 0.536 & 0.666 & 0.184 & 0.628 & 0.362 \\
                        & \pmS{0.032} & \pmS{0.026} & \pmS{0.021} & \pmS{0.027} & \pmS{0.013} & \pmS{0.018} & \pmS{0.005} & \pmS{0.025} & \pmS{0.007} \\
\multirow{2}{*}{MoCo}   & 0.280 & 0.791 & 0.484 & 0.791 & 0.603 & 0.716 & 0.148 & 0.678 & 0.360 \\
                        & \pmS{0.050} & \pmS{0.020} & \pmS{0.034} & \pmS{0.014} & \pmS{0.014} & \pmS{0.008} & \pmS{0.000} & \pmS{0.004} & \pmS{0.001} \\
\multirow{2}{*}{SimSiam}& 0.154 & 0.673 & 0.362 & 0.585 & 0.487 & 0.546 & 0.100 & 0.654 & 0.322 \\
                        & \pmS{0.043} & \pmS{0.033} & \pmS{0.036} & \pmS{0.043} & \pmS{0.055} & \pmS{0.038} & \pmS{0.019} & \pmS{0.055} & \pmS{0.011} \\
\multirow{2}{*}{NNCLR}  & 0.332 & 0.665 & 0.465 & 0.695 & 0.500 & 0.617 & 0.147 & 0.662 & 0.353 \\
                        & \pmS{0.084} & \pmS{0.035} & \pmS{0.060} & \pmS{0.019} & \pmS{0.008} & \pmS{0.010} & \pmS{0.002} & \pmS{0.019} & \pmS{0.008} \\
\multirow{2}{*}{BYOL}   & 0.000 & 0.694 & 0.277 & 0.582 & 0.707 & 0.632 & 0.028 & 0.705 & 0.299 \\
                        & \pmS{0.009} & \pmS{0.037} & \pmS{0.016} & \pmS{0.021} & \pmS{0.027} & \pmS{0.022} & \pmS{0.003} & \pmS{0.003} & \pmS{0.001} \\
\multirow{2}{*}{VICReg} & 0.515 & 0.709 & 0.592 & \textbf{0.830} & 0.581 & \textbf{0.730} & 0.208 & 0.622 & 0.374 \\
                        & \pmS{0.018} & \pmS{0.014} & \pmS{0.011} & \pmS{0.024} & \pmS{0.013} & \pmS{0.016} & \pmS{0.008} & \pmS{0.003} & \pmS{0.004} \\
Barlow                  & 0.471 & 0.627 & 0.533 & 0.784 & 0.533 & 0.684 & 0.173 & 0.651 & 0.364 \\
Twins                   & \pmS{0.032} & \pmS{0.029} & \pmS{0.026} & \pmS{0.011} & \pmS{0.004} & \pmS{0.008} & \pmS{0.007} & \pmS{0.010} & \pmS{0.000}\\
\midrule
\multirow{2}{*}{Concerto} & 0.497 & 0.559 & 0.522 & 0.000 & 0.627 & 0.251 & 0.152 & 0.711 & 0.376 \\
                        & \pmS{0.036} & \pmS{0.013} & \pmS{0.027} & \pmS{0.000} & \pmS{0.004} & \pmS{0.001} & \pmS{0.008} & \pmS{0.008} & \pmS{0.008} \\

\multirow{2}{*}{CLEAR} & \textbf{0.684} & 0.385 & 0.565 & 0.615 & 0.262 & 0.474 & \textbf{0.876} & 0.326 & 0.656 \\
& \pmS{0.042} & \pmS{0.008} & \pmS{0.027} & \pmS{0.004} & \pmS{0.007} & \pmS{0.005} & \pmS{0.088} & \pmS{0.049} & \pmS{0.050} \\

\multirow{2}{*}{CLAIRE} & 0.467 & \textbf{0.978} & \textbf{0.672} & 0.511 & \textbf{0.974} & 0.696 & 0.315 & \textbf{0.907} & 0.552 \\
                        & \pmS{0.015} & \pmS{0.007} & \pmS{0.010} & \pmS{0.024} & \pmS{0.000} & \pmS{0.014} & \pmS{0.324} & \pmS{0.058} & \pmS{0.176} \\
\midrule
\multirow{2}{*}{scVI} & 0.723 & 0.254 & 0.536 & 0.768 & 0.659 & 0.725 & 0.716 & 0.659 & \textbf{0.693} \\
                        & \pmS{0.021} & \pmS{0.017} & \pmS{0.019} & \pmS{0.007} & \pmS{0.003} & \pmS{0.006} & \pmS{0.176} & \pmS{0.017} & \pmS{0.111} \\
\midrule
\multirow{2}{*}{PCA} & 0.538 & 0.297 & 0.442 & 0.635 & 0.164 & 0.446 & 0.748 & 0.245 & 0.547 \\
                        & \pmS{0.027} & \pmS{0.000} & \pmS{0.016} & \pmS{0.014} & \pmS{0.003} & \pmS{0.009} & \pmS{0.114} & \pmS{0.062} & \pmS{0.093} \\
\bottomrule
\end{tabular}
\label{tab:integration_supplement}
\end{table}

\begin{table*}[ht!]
\centering
\caption{Uni-modal cell-typing with CLEAR augmentations. We define one technology (10X 5' v2) of the Immune Cell Atlas as a hold-out set, train the encoder and knn-classifier. The generic model VICReg outperforms all other methods.}
\vspace{0.3cm}
\setlength\tabcolsep{20pt}
\begin{tabular}{l|cc}
\toprule
\multirow{2}{*}{\textbf{Method}} & \multicolumn{2}{c}{\textbf{Immune Cell Atlas (10X 5' v2)}} \\
& Macro F1 & Acc\\
\midrule
SimCLR     & 0.788 \pmS{0.004} & 0.830 \pmS{0.003}\\
MoCo       & 0.794 \pmS{0.006} & 0.835 \pmS{0.014} \\
SimSiam    & 0.711 \pmS{0.015} & 0.768 \pmS{0.007}\\
NNCLR      & 0.740 \pmS{0.010} & 0.804 \pmS{0.007}\\
BYOL       & 0.680 \pmS{0.002} & 0.724  \pmS{0.009}\\
VICReg     & 0.820 \pmS{0.012} & 0.866 \pmS{0.003} \\
Barlow Twins                    & 0.727 \pmS{0.004} & 0.752 \pmS{0.007}\\
\midrule

CLEAR   & 0.806  \pmS{0.000} & 0.855  \pmS{0.000} \\
CLAIRE  & 0.436 \pmS{0.000} & 0.492 \pmS{0.002} \\
\midrule

scGPT (zero-shot) & 0.358 \pmS{0.004} & 0.439\pmS{0.001} \\
scGPT (finetuned)& \textbf{0.835}\pmS{0.008} & 0.850\pmS{0.008} \\
Geneformer (finetuned) & 0.831 \pmS{0.000} & \textbf{0.878} \pmS{0.000} \\
scBERT & 0.818 \pmS{0.009} & 0.873 \pmS{0.004} \\
scVI  & 0.750 \pmS{0.001} & 0.804 \pmS{0.000} \\

\midrule
SCDC  & 0.595 \pmS{0.009} & 0.642 \pmS{0.000} \\
PCA  & 0.071 \pmS{0.007} & 0.124  \pmS{0.000} \\

\bottomrule
\end{tabular}
\label{tab:qr_big_transposed}
\end{table*}

\begin{table}[t!]
\vspace{-1cm}
\centering
\caption{Cell-type annotation with CLEAR augmentations on the Pancreas dataset. We define individual studies as holdout sets during training. Accuracy and Macro F1 are computed on the holdout set.}
\setlength\tabcolsep{10pt}
\begin{tabular}{l|cc|cc|cc|cc}
\toprule
\multirow{2}{*}{\textbf{Method}} & \multicolumn{2}{c|}{\textbf{Mutaro et al.}} & \multicolumn{2}{c|}{\textbf{Segerstolpe et al.}} & \multicolumn{2}{c|}{\textbf{Wang et al.}} & \multicolumn{2}{c}{\textbf{Xin et al.}} \\
& Macro F1 & Acc & Macro F1 & Acc & Macro F1 & Acc & Macro F1 & Acc  \\
\midrule
\multirow{2}{*}{SimCLR}     &0.796&0.936&0.781&0.946&\textbf{0.894}&\textbf{0.945}&0.778&0.819 \\
                            & \pmS{0.014} & \pmS{0.012} & \pmS{0.022} & \pmS{0.011} & \pmS{0.005} & \pmS{0.010} & \pmS{0.050} & \pmS{0.019} \\
\multirow{2}{*}{MoCo}       &0.825&0.939&0.799&0.936&0.844&0.919&0.722&0.798 \\
                            & \pmS{0.045} & \pmS{0.013} & \pmS{0.033} & \pmS{0.006} & \pmS{0.020} & \pmS{0.021} & \pmS{0.045} & \pmS{0.010} \\
\multirow{2}{*}{SimSiam}    &0.679&0.875&0.595&0.803&0.586&0.798&0.464&0.718 \\
                            & \pmS{0.042} & \pmS{0.026} & \pmS{0.022} & \pmS{0.014} & \pmS{0.036} & \pmS{0.022} & \pmS{0.071} & \pmS{0.037} \\
\multirow{2}{*}{NNCLR}      &0.707&0.897&0.619&0.854&0.729&0.868&0.472&0.721 \\
                            & \pmS{0.043} & \pmS{0.024} & \pmS{0.013} & \pmS{0.012} & \pmS{0.019} & \pmS{0.012} & \pmS{0.033} & \pmS{0.025} \\
\multirow{2}{*}{BYOL}       &0.720&0.884&0.670&0.854&0.656&0.813&0.524&0.724 \\
                            & \pmS{0.037} & \pmS{0.021} & \pmS{0.005} & \pmS{0.006} & \pmS{0.048} & \pmS{0.008} & \pmS{0.062} & \pmS{0.013} \\
\multirow{2}{*}{VICReg}     &0.853&0.947&0.855&\textbf{0.976}&0.877&0.937&0.830&0.839 \\
                            & \pmS{0.039} & \pmS{0.002} & \pmS{0.008} & \pmS{0.006} & \pmS{0.008} & \pmS{0.004} & \pmS{0.023} & \pmS{0.010} \\
Barlow                      &0.700&0.868&0.673&0.870&0.737&0.878&0.487&0.720\\
Twins                       & \pmS{0.010} & \pmS{0.005} & \pmS{0.017} & \pmS{0.007} & \pmS{0.012} & \pmS{0.010} & \pmS{0.003} & \pmS{0.009} \\
\midrule
\multirow{2}{*}{Concerto}   & 0.106 & 0.431 & 0.113 & 0.419 & 0.105 & 0.435 & 0.112 & 0.406 \\
& \pmS{0.000} & \pmS{0.000} & \pmS{0.000} & \pmS{0.000} & \pmS{0.000} & \pmS{0.000} & \pmS{0.000} & \pmS{0.000} \\

\multirow{2}{*}{CLEAR}   & \textbf{0.950} & \textbf{0.961} & 0.898 & 0.967 & 0.891 & 0.941 & 0.987 & 0.994 \\
& \pmS{0.000} & \pmS{0.001} & \pmS{0.000} & \pmS{0.002} & \pmS{0.000} & \pmS{0.007} & \pmS{0.000} & \pmS{0.001} \\

\multirow{2}{*}{CLAIRE}  & 0.941 & 0.937 & \textbf{0.919} & 0.955 & 0.893 & 0.945& 0.965 & 0.989 
\\
& \pmS{0.000} & \pmS{0.002} & \pmS{0.000} & \pmS{0.002} & \pmS{0.000} & \pmS{0.001} & \pmS{0.000} & \pmS{0.002} \\

\midrule

scGPT  & 0.502 & 0.765 & 0.549 &0.826 & 0.327 & 0.519 & 0.581 & 0.786  \\
(zero-shot)&\pmS{0.00}&\pmS{0.00}&\pmS{0.02}&\pmS{0.01}&\pmS{0.15}&\pmS{0.21}&\pmS{0.00}&\pmS{0.00} \\

scGPT  & 0.850 & 0.917 & 0.803 & 0.955 & 0.466 & 0.560 & \textbf{0.989} & \textbf{0.995} \\
(finetuned)&\pmS{0.001}&\pmS{0.003}&\pmS{0.02}&\pmS{0.010}&\pmS{0.154}&\pmS{0.217}&\pmS{0.006}&\pmS{0.002} \\

Geneformer & 0.622 & 0.916 & 0.630 & 0.944 & 0.673 & 0.891& 0.564 & 0.993
\\
(finetuned)& \pmS{0.000} & \pmS{0.000} & \pmS{0.000} & \pmS{0.000} & \pmS{0.000} & \pmS{0.000} & \pmS{0.000} & \pmS{0.000} \\

\multirow{2}{*}{scBERT}  & 0.642 & 0.919 & 0.715 & 0.953 & 0.710 & 0.902& 0.975 & 0.990
\\
& \pmS{0.004} & \pmS{0.012} & \pmS{0.002} & \pmS{0.002} & \pmS{0.0830} & \pmS{0.001} & \pmS{0.007} & \pmS{0.000} \\

\multirow{2}{*}{scVI}  & 0.616 & 0.889 & 0.635 & 0.903 & 0.680 & 0.897 & 0.741 & 0.916 \\
& \pmS{0.015} & \pmS{0.012} & \pmS{0.127} & \pmS{0.023} & \pmS{0.135} & \pmS{0.023} & \pmS{0.171} & \pmS{0.023} \\

\midrule

\multirow{2}{*}{SCDC}  & 0.571 & 0.860 & 0.581 & 0.875 & 0.689 & 0.908& 0.442 & 0.800
\\
& \pmS{0.004} & \pmS{0.021} & \pmS{0.001} & \pmS{0.015} & \pmS{0.003} & \pmS{0.002} & \pmS{0.002} & \pmS{0.009} \\

\multirow{2}{*}{PCA}   & 0.071 & 0.082 & 0.171 & 0.236 & 0.195 & 0.371 & 0.054 & 0.427 \\
& \pmS{0.000} & \pmS{0.00} & \pmS{0.000} & \pmS{0.000} & \pmS{0.000} & \pmS{0.000} & \pmS{0.000} & \pmS{0.000} \\

\bottomrule
\end{tabular}
\label{tab:template_qr}
\vspace{-1cm}
\end{table}

\begin{table*}[t!]
\centering
\caption{Cell-type annotation for multi-modal datasets with CLEAR pipeline. On the left, two modalities (RNA + Protein or GEX (gene expression) + ADT (protein abundance)) were used during inference. On the right, we show inference performance with a single modality (RNA or GEX). All models were trained with two modalities.}
\vspace{0.3cm}
\setlength\tabcolsep{9pt}
\begin{tabular}{l|cc|cc||cc|cc}
\toprule
\multirow{3}{*}{\textbf{Method}} & \multicolumn{2}{c|}{\textbf{RNA + Protein}} & \multicolumn{2}{c||}{\textbf{GEX + ADT}} & \multicolumn{2}{c|}{\textbf{RNA}} & \multicolumn{2}{c}{\textbf{GEX}}\\
\cmidrule{2-9}
 & \multicolumn{2}{c|}{\textbf{PBMC-M}} & \multicolumn{2}{c||}{\textbf{BMMC}} & \multicolumn{2}{c|}{\textbf{PBMC-M}} & \multicolumn{2}{c}{\textbf{BMMC}}\\
& Macro F1 & Acc & Macro F1 & Acc & Macro F1 & Acc & Macro F1 & Acc  \\
\midrule

\multirow{2}{*}{SimCLR}  & \textbf{0.950} & \textbf{0.977} & 0.770 & 0.876 & 0.906 & 0.946 & 0.749 & 0.848 \\
                         & \pmS{0.002} & \pmS{0.001}	& \pmS{0.028} & \pmS{0.022}	& \pmS{0.002} & \pmS{0.001}	& \pmS{0.050} & \pmS{0.035} \\
\multirow{2}{*}{MoCo}    & 0.930 & 0.969 & 0.609 & 0.771 & 0.778 & 0.835 & 0.630 & 0.717 \\
                         & \pmS{0.007} & \pmS{0.004}	& \pmS{0.001} & \pmS{0.041}	& \pmS{0.016} & \pmS{0.010}	& \pmS{0.065} & \pmS{0.073} \\
\multirow{2}{*}{SimSiam} & 0.933 & 0.968 & 0.666 & 0.820 & 0.846 & 0.884 & 0.670 & 0.792 \\
                         & \pmS{0.002} & \pmS{0.001}	& \pmS{0.069} & \pmS{0.036}	& \pmS{0.016} & \pmS{0.019}	& \pmS{0.082} & \pmS{0.056} \\
\multirow{2}{*}{NNCLR}   & 0.941 & 0.971 & 0.734 & 0.856 & 0.860 & 0.901 & 0.703 & 0.806 \\
                         & \pmS{0.004} & \pmS{0.002}	& \pmS{0.055} & \pmS{0.043}	& \pmS{0.008} & \pmS{0.003}	& \pmS{0.085} & \pmS{0.070} \\
\multirow{2}{*}{BYOL}    & 0.933 & 0.968 & 0.737 & 0.847 & 0.857 & 0.898 & 0.704 & 0.795 \\
                         & \pmS{0.008} & \pmS{0.004}	& \pmS{0.051} & \pmS{0.037}	& \pmS{0.047} & \pmS{0.034}	& \pmS{0.048} & \pmS{0.054} \\
\multirow{2}{*}{VICReg}  & \textbf{0.950} & \textbf{0.977} & 0.808 & 0.899 & 0.923  &0.957 & 0.785 & 0.887 \\
                         & \pmS{0.001} & \pmS{0.000}	& \pmS{0.021} & \pmS{0.017}	& \pmS{0.005} & \pmS{0.004}	& \pmS{0.014} & \pmS{0.015} \\
Barlow                   & 0.949 & 0.976 & 0.766 & 0.863 & 0.890 & 0.919 & 0.733 & 0.816 \\
Twins                    & \pmS{0.001} & \pmS{0.001}	& \pmS{0.012} & \pmS{0.003}	& \pmS{0.014} & \pmS{0.008}	& \pmS{0.013} & \pmS{0.017} \\
\midrule
\multirow{2}{*}{Concerto} & 0.892 & 0.947 & 0.673 & 0.825 & \multirow{2}{*}{---} & \multirow{2}{*}{---} & \multirow{2}{*}{---} & \multirow{2}{*}{---} \\
& \pmS{0.000} & \pmS{0.001} & \pmS{0.000}   & \pmS{0.000} & & & &  \\

\multirow{2}{*}{scCLIP} & 0.699 & 0.851 & 0.557 & 0.797 & 0.728 & 0.857 & 0.635 & 0.818 \\
& \pmS{0.009} & \pmS{0.025} & \pmS{0.015}   & \pmS{0.008} & \pmS{0.019} & \pmS{0.004} & \pmS{0.071}   & \pmS{0.051} \\

\midrule

\multirow{2}{*}{scButterfly} & 0.949 & 0.976 & \textbf{0.844} & \textbf{0.924} & \textbf{0.946} & \textbf{0.973} & \textbf{0.831} & \textbf{0.920} \\
& \pmS{0.000} & \pmS{0.000} & \pmS{0.002}   & \pmS{0.000} & \pmS{0.000} & \pmS{0.000} & \pmS{0.000}   & \pmS{0.001} \\

\multirow{2}{*}{scTEL} & 0.173 & 0.211 & 0.039 & 0.149 & --- & --- & ---- & --- \\
& \pmS{0.001} & \pmS{0.031} & \pmS{0.003}   & \pmS{0.004} &  &  &    &  \\

\midrule

\multirow{2}{*}{totalVI} & 0.829 & \textbf{0.977} & 0.829 & 0.911 & \multirow{2}{*}{---} & \multirow{2}{*}{---} & \multirow{2}{*}{---} & \multirow{2}{*}{---} \\
& \pmS{0.021} & \pmS{0.0158} & \pmS{0.023}   & \pmS{0.015} &  &  &  & \\

\bottomrule
\end{tabular}

\label{tab:multimodal_qr}
\end{table*}

\begin{table}[t!]
\centering
\caption{Missing modality prediction for methods trained with the CLEAR pipeline on multi-modal datasets. We show the average Pearson correlation between the original and inferred missing modality: protein for PBMC-M, and ADT (protein abundance) for BMMC.}
\vspace{0.3cm}
\setlength\tabcolsep{20pt}
\begin{tabular}{l|c|c}
\toprule
\multirow{2}{*}{\textbf{Method}} & \multicolumn{1}{c|}{\textbf{PBMC-M}} & \multicolumn{1}{c}{\textbf{BMMC}} \\
& Pearson Mean & Pearson Mean \\
\midrule

SimCLR  & \textbf{0.866} \pmS{0.001}  &  0.757 \pmS{0.002} \\
MoCo    &      0.856 \pmS{0.001}  &  0.721 \pmS{0.004} \\
                         
SimSiam &      0.859 \pmS{0.002}&      0.748 \pmS{0.002} \\
NNCLR   &      0.861  \pmS{0.002} &      0.751 \pmS{0.001} \\
BYOL    &      0.860 \pmS{0.000}  &      0.738  \pmS{0.002}\\
VICReg  &      0.865  \pmS{0.001}&      \textbf{0.759} \pmS{0.001} \\
Barlow   Twins                &      0.864 \pmS{0.001} &      0.755 \pmS{0.001}  \\
\midrule
Concerto &      0.742 \pmS{0.006} &  0.542  \pmS{0.001}\\
scCLIP & 0.614 \pmS{0.003} &  0.175  \pmS{0.005}\\
\midrule
scButterfly  (kNN) & 0.856 \pmS{0.000} &  0.651  \pmS{0.001}\\
scButterfly (generated) & 0.840 \pmS{0.000} &  0.624  \pmS{0.002}\\
scTEL (100 epochs) & 0.022 \pmS{0.005} &  0.047  \pmS{0.002}\\
\bottomrule
\end{tabular}

\label{tab:multimodal_mp}
\end{table}

\begin{table*}[!ht]
\centering
\caption{Batch correction results for methods with and without projection layer during inference. All methods are trained using the CLEAR pipeline. Results are not min-max scaled for easier comparison.}
\vspace{0.3cm}
\setlength\tabcolsep{5pt}
\begin{tabular}{l|ccc|ccc||ccc|ccc}
\toprule
&\multicolumn{6}{c||}{\textbf{Encoder}} & \multicolumn{6}{c}{\textbf{Encoder + Projection}} \\
\multirow{2}{*}{\textbf{Method}} & \multicolumn{3}{c|}{\textbf{MCA}} & \multicolumn{3}{c||}{\textbf{HIC}} & \multicolumn{3}{c|}{\textbf{MCA}} & \multicolumn{3}{c}{\textbf{HIC}} \\
& Bio & Batch & Total & Bio & Batch & Total & Bio & Batch & Total & Bio & Batch & Total\\
\midrule
\multirow{2}{*}{SimCLR}& 0.620 & 0.633 & 0.625 & 0.683 & 0.567 & 0.637 & 0.575 & 0.658 & 0.608 & 0.674 & 0.575 & 0.635 \\
& \pmS{0.020} & \pmS{0.007} & \pmS{0.014} & \pmS{0.008} & \pmS{0.001} & \pmS{0.004} & \pmS{0.003} & \pmS{0.006} & \pmS{0.000} & \pmS{0.010} & \pmS{0.009} & \pmS{0.002} \\

\multirow{2}{*}{MoCo} & 0.561 & 0.706 & 0.619 & \textbf{0.721} & 0.585 & 0.667 & 0.536 & \textbf{0.726} & 0.612 & 0.705 & 0.602 & 0.664 \\
& \pmS{0.013} & \pmS{0.002} & \pmS{0.007} & \pmS{0.000} & \pmS{0.003} & \pmS{0.001} & \pmS{0.000} & \pmS{0.015} & \pmS{0.006} & \pmS{0.019} & \pmS{0.001} & \pmS{0.012} \\

\multirow{2}{*}{SimSiam}& 0.519 & 0.656 & 0.574 & 0.623 & 0.545 & 0.592 & 0.435 & 0.681 & 0.533 & 0.598 & 0.540 & 0.575 \\
& \pmS{0.008} & \pmS{0.025} & \pmS{0.006} & \pmS{0.040} & \pmS{0.012} & \pmS{0.029} & \pmS{0.009} & \pmS{0.027} & \pmS{0.006} & \pmS{0.038} & \pmS{0.017} & \pmS{0.030} \\

\multirow{2}{*}{NNCLR}  & 0.570 & 0.646 & 0.600 & 0.680 & 0.550 & 0.628 & 0.493 & 0.675 & 0.566 & 0.657 & 0.557 & 0.617 \\
& \pmS{0.012} & \pmS{0.003} & \pmS{0.008} & \pmS{0.023} & \pmS{0.003} & \pmS{0.013} & \pmS{0.001} & \pmS{0.013} & \pmS{0.004} & \pmS{0.026} & \pmS{0.005} & \pmS{0.018} \\

\multirow{2}{*}{BYOL} & 0.486 & 0.670 & 0.560 & 0.637 & 0.602 & 0.623 & 0.398 & 0.667 & 0.506 & 0.552 & \textbf{0.607} & 0.574 \\
& \pmS{0.017} & \pmS{0.003} & \pmS{0.012} & \pmS{0.007} & \pmS{0.000} & \pmS{0.004} & \pmS{0.004} & \pmS{0.028} & \pmS{0.013} & \pmS{0.010} & \pmS{0.007} & \pmS{0.009} \\

\multirow{2}{*}{VICReg} & 0.624 & 0.653 & 0.636 & 0.716 & 0.584 & 0.663 & 0.587 & 0.699 & \textbf{0.632} & \textbf{0.726} & 0.588 & \textbf{0.671} \\
& \pmS{0.005} & \pmS{0.009} & \pmS{0.006} & \pmS{0.007} & \pmS{0.002} & \pmS{0.003} & \pmS{0.002} & \pmS{0.009} & \pmS{0.005} & \pmS{0.005} & \pmS{0.005} & \pmS{0.005} \\

Barlow    & 0.606 & 0.623 & 0.613 & 0.716 & 0.566 & 0.656 & 0.544 & 0.664 & 0.592 & 0.675 & 0.562 & 0.630 \\
Twins  & \pmS{0.009} & \pmS{0.013} & \pmS{0.000} & \pmS{0.006} & \pmS{0.006} & \pmS{0.001} & \pmS{0.018} & \pmS{0.004} & \pmS{0.012} & \pmS{0.029} & \pmS{0.006} & \pmS{0.020} \\

\midrule
\multirow{2}{*}{Concerto} & \multirow{2}{*}{---} & \multirow{2}{*}{---}  & \multirow{2}{*}{---}  & \multirow{2}{*}{---}  & \multirow{2}{*}{---}   & \multirow{2}{*}{---}    & 0.634 & 0.529 & 0.592 & 0.357 & 0.470 & 0.402  \\
&     &     &     &     &     &     & \pmS{0.002} & \pmS{0.000} & \pmS{0.001} & \pmS{0.000} & \pmS{0.001} & \pmS{0.001}\\

\multirow{2}{*}{CLEAR} & \textbf{0.696} & 0.438 & 0.593 & 0.642 & 0.408 & 0.549 & \multirow{2}{*}{---} & \multirow{2}{*}{---}  & \multirow{2}{*}{---}  & \multirow{2}{*}{---}  & \multirow{2}{*}{---}   & \multirow{2}{*}{---}   \\
& \pmS{0.010} & \pmS{0.001} & \pmS{0.006} & \pmS{0.003} & \pmS{0.006} & \pmS{0.001} &     &     &     &     &     &      \\

\multirow{2}{*}{CLAIRE}  & 0.689 & \textbf{0.763} & \textbf{0.718} & 0.699 & \textbf{0.700} & \textbf{0.699} & \multirow{2}{*}{---} & \multirow{2}{*}{---}  & \multirow{2}{*}{---}  & \multirow{2}{*}{---}  & \multirow{2}{*}{---}   & \multirow{2}{*}{---} \\
& \pmS{0.018} & \pmS{0.008} & \pmS{0.014} & \pmS{0.029} & \pmS{0.003} & \pmS{0.018}  &     &     &     &     &     &  \\
\midrule
\multirow{2}{*}{PCA} & 0.651 & 0.348 & 0.530 & 0.654 & 0.320 & 0.521 & \textbf{0.651} & 0.348 & 0.530 & 0.654 & 0.302 & 0.521 \\
& \pmS{0.001} & \pmS{0.000} & \pmS{0.000} & \pmS{0.008} & \pmS{0.000} & \pmS{0.005} & \pmS{0.001} & \pmS{0.000} & \pmS{0.000} & \pmS{0.008} & \pmS{0.000} & \pmS{0.005} \\
\bottomrule
\end{tabular}
\label{tab:unimodal_projection}
\vspace{-1cm}
\end{table*}

\begin{table}[t!]
\centering
\caption{Data integration for methods using the CLEAR pipeline on multimodal datasets. We compare the effect of retaining the projection head during inference to the representation quality when using only the encoder. This table is not min-max scaled.}
\vspace{0.3cm}
\setlength\tabcolsep{5pt}
\begin{tabular}{l|ccc|ccc||ccc|ccc}
\toprule
&\multicolumn{6}{c||}{\textbf{Encoder}} & \multicolumn{6}{c}{\textbf{Encoder + Projection}} \\
\multirow{3}{*}{\textbf{Method}} & \multicolumn{3}{c|}{\textbf{PBMC-M}} & \multicolumn{3}{c||}{\textbf{BMMC}} & \multicolumn{3}{c|}{\textbf{PBMC-M}} & \multicolumn{3}{c}{\textbf{BMMC}} \\
& Bio & Batch & Total & Bio & Batch & Total & Bio & Batch & Total & Bio & Batch & Total \\
\midrule

\multirow{2}{*}{SimCLR} & 0.738 & 0.518 & 0.650 & 0.718 & 0.574 & 0.660 & 0.741 & 0.513 & 0.650 & 0.712 & 0.567 & 0.654 \\
& \pmS{0.000} & \pmS{0.010} & \pmS{0.004} & \pmS{0.016} & \pmS{0.001} & \pmS{0.010} & \pmS{0.021} & \pmS{0.005} & \pmS{0.015} & \pmS{0.004} & \pmS{0.001} & \pmS{0.002} \\
\multirow{2}{*}{MoCo}   & 0.776 & \textbf{0.570} & 0.694 & 0.677 & \textbf{0.582} & 0.639 & 0.762 & \textbf{0.606} & \textbf{0.700} & 0.592 & \textbf{0.637} & 0.610\\
& \pmS{0.001} & \pmS{0.007} & \pmS{0.002} & \pmS{0.021} & \pmS{0.004} & \pmS{0.014} & \pmS{0.009} & \pmS{0.004} & \pmS{0.004} & \pmS{0.025} & \pmS{0.007} & \pmS{0.012} \\
\multirow{2}{*}{SimSiam}& 0.766 & 0.563 & 0.685 & 0.677 & 0.559 & 0.629 & 0.767 & 0.565 & 0.686 & 0.635 & 0.553 & 0.602\\
& \pmS{0.051} & \pmS{0.012} & \pmS{0.036} & \pmS{0.011} & \pmS{0.000} & \pmS{0.006} & \pmS{0.024} & \pmS{0.006} & \pmS{0.017} & \pmS{0.021} & \pmS{0.002} & \pmS{0.014}\\
\multirow{2}{*}{NNCLR}  & 0.750 & 0.535 & 0.664 & 0.706 & 0.565 & 0.650 & 0.743 & 0.533 & 0.659 & 0.695 & 0.575 & 0.647 \\
& \pmS{0.007} & \pmS{0.010} & \pmS{0.000} & \pmS{0.002} & \pmS{0.001} & \pmS{0.002} & \pmS{0.032} & \pmS{0.024} & \pmS{0.029} & \pmS{0.014} & \pmS{0.006} & \pmS{0.011}\\
\multirow{2}{*}{BYOL}   & \textbf{0.789} & 0.557 & \textbf{0.696} & 0.701 & 0.574 & 0.651 & \textbf{0.783} & 0.543 & 0.687 & 0.691 & 0.584 & 0.648 \\
& \pmS{0.015} & \pmS{0.000} & \pmS{0.009} & \pmS{0.005} & \pmS{0.007} & \pmS{0.006} & \pmS{0.007} & \pmS{0.009} & \pmS{0.008} & \pmS{0.006} & \pmS{0.002} & \pmS{0.004}\\
\multirow{2}{*}{VICReg} & 0.749 & 0.478 & 0.641 & \textbf{0.722} & \textbf{0.582} & 0.666 & 0.763 & 0.491 & 0.654 & \textbf{0.714} & 0.578 & \textbf{0.660} \\
& \pmS{0.008} & \pmS{0.017} & \pmS{0.002} & \pmS{0.002} & \pmS{0.001} & \pmS{0.000} & \pmS{0.001} & \pmS{0.003} & \pmS{0.002} & \pmS{0.000} & \pmS{0.001} & \pmS{0.001}\\
Barlow                  & 0.755 & 0.509 & 0.657 & 0.704 & 0.577 & 0.653 & 0.712 & 0.506 & 0.630 & 0.704 & 0.581 & 0.655 \\
Twins                   & \pmS{0.018} & \pmS{0.015} & \pmS{0.017} & \pmS{0.001} & \pmS{0.005} & \pmS{0.003} & \pmS{0.006} & \pmS{0.003} & \pmS{0.005} & \pmS{0.002} & \pmS{0.009} & \pmS{0.002}\\
\midrule
\multirow{2}{*}{Concerto} & --- &  ---  & --- &  --- & --- &  --- & 0.773 & 0.436 & 0.638 & 0.604 & 0.525 & 0.573 \\
                          & --- & --- & --- & --- & --- & --- &\pmS{0.117} &\pmS{0.006} &\pmS{0.072} &\pmS{0.089} &\pmS{0.01 } &\pmS{0.054}\\

\midrule
\multirow{2}{*}{PCA} & 0.602 & 0.504 & 0.563 & 0.558 & 0.322 & 0.464 & 0.602 & 0.504 & 0.563 & 0.558 & 0.322 & 0.464 \\
                          &\pmS{0.000} &\pmS{0.000} &\pmS{0.000} &\pmS{0.000} &\pmS{0.000} &\pmS{0.000} &\pmS{0.000} &\pmS{0.000} &\pmS{0.000} &\pmS{0.000} &\pmS{0.000} &\pmS{0.000} \\

\bottomrule
\end{tabular}

\label{tab:multimodal_projection_bc}
\end{table}

\begin{table*}[ht!]
\centering

\caption{Batch correction benchmark for methods trained using the CLEAR pipeline with domain specific batch normalization (DSBN). Results are not min-max scaled for easier comparison.}
\setlength\tabcolsep{5pt}
\begin{tabular}{l|ccc|ccc||ccc|ccc}
\toprule
&\multicolumn{6}{c||}{\textbf{Batch Normalization}} & \multicolumn{6}{c}{\textbf{DSBN}} \\
\multirow{2}{*}{\textbf{Method}} & \multicolumn{3}{c|}{\textbf{HIC}} & \multicolumn{3}{c||}{\textbf{MCA}} & \multicolumn{3}{c|}{\textbf{HIC}} & \multicolumn{3}{c}{\textbf{MCA}} \\
& Bio & Batch & Total & Bio & Batch & Total & Bio & Batch & Total & Bio & Batch & Total\\
\midrule
\multirow{2}{*}{SimCLR} & 0.703 & 0.573 & 0.651 & 0.627 & 0.644 & 0.634 & 0.680 & 0.583 & 0.641 & 0.624 & 0.636 & 0.629 \\
& \pmS{0.022} & \pmS{0.014} & \pmS{0.009} & \pmS{0.009} & \pmS{0.026} & \pmS{0.014} & \pmS{0.020} & \pmS{0.015} & \pmS{0.008} & \pmS{0.008} & \pmS{0.009} & \pmS{0.006}\\
\multirow{2}{*}{MoCo} & 0.707 & 0.582 & 0.657 & 0.518 & \textbf{0.731} & 0.603 & 0.648 & \textbf{0.612} & 0.633 & 0.549 & \textbf{0.697} & 0.608 \\
& \pmS{0.010} & \pmS{0.020} & \pmS{0.011} & \pmS{0.060} & \pmS{0.032} & \pmS{0.047} & \pmS{0.040} & \pmS{0.004} & \pmS{0.022} & \pmS{0.013} & \pmS{0.003} & \pmS{0.008} \\
\multirow{2}{*}{SimSiam}& 0.619 & 0.544 & 0.589 & 0.523 & 0.668 & 0.581 & 0.603 & 0.595 & 0.600 & 0.502 & 0.635 & 0.555 \\
& \pmS{0.053} & \pmS{0.043} & \pmS{0.049} & \pmS{0.069} & \pmS{0.072} & \pmS{0.069} & \pmS{0.074} & \pmS{0.025} & \pmS{0.034} & \pmS{0.018} & \pmS{0.013} & \pmS{0.015} \\
\multirow{2}{*}{NNCLR}  & 0.658 & 0.546 & 0.613 & 0.574 & 0.637 & 0.599 & 0.659 & 0.590 & 0.632 & 0.543 & 0.651 & 0.587 \\
& \pmS{0.015} & \pmS{0.011} & \pmS{0.010} & \pmS{0.120} & \pmS{0.056} & \pmS{0.087} & \pmS{0.011} & \pmS{0.006} & \pmS{0.009} & \pmS{0.019} & \pmS{0.001} & \pmS{0.011} \\
\multirow{2}{*}{BYOL}  & 0.607 & \textbf{0.624} & 0.614 & 0.483 & 0.679 & 0.561 & 0.576 & 0.600 & 0.586 & 0.473 & 0.673 & 0.553 \\
& \pmS{0.024} & \pmS{0.016} & \pmS{0.020} & \pmS{0.005} & \pmS{0.042} & \pmS{0.019} & \pmS{0.050} & \pmS{0.018} & \pmS{0.023} & \pmS{0.021} & \pmS{0.014} & \pmS{0.017} \\
\multirow{2}{*}{VICReg} & 0.706 & 0.592 & \textbf{0.661} & 0.615 & 0.665 & \textbf{0.635} & 0.674 & 0.591 & 0.641 & 0.619 & 0.649 & \textbf{0.631} \\
& \pmS{0.034} & \pmS{0.019} & \pmS{0.017} & \pmS{0.016} & \pmS{0.020} & \pmS{0.017} & \pmS{0.056} & \pmS{0.012} & \pmS{0.033} & \pmS{0.004} & \pmS{0.002} & \pmS{0.003} \\
Barlow                  & \textbf{0.713} & 0.572 & 0.656 & 0.603 & 0.636 & 0.617 & \textbf{0.707} & 0.577 & \textbf{0.655} & 0.603 & 0.634 & 0.615 \\
Twins                  & \pmS{0.014} & \pmS{0.008} & \pmS{0.005} & \pmS{0.010} & \pmS{0.064} & \pmS{0.027} & \pmS{0.004} & \pmS{0.003} & \pmS{0.002} & \pmS{0.020} & \pmS{0.011} & \pmS{0.008} \\
\midrule
\multirow{2}{*}{Concerto} & 0.357 & 0.470 & 0.402 & 0.635 & 0.529 & 0.593 & \multirow{2}{*}{---} & \multirow{2}{*}{---} & \multirow{2}{*}{---} & \multirow{2}{*}{---} & \multirow{2}{*}{---} & \multirow{2}{*}{---} \\
& \pmS{0.000} & \pmS{0.029} & \pmS{0.012} & \pmS{0.016} & \pmS{0.028} & \pmS{0.021} & & & & &  & \\
\midrule
\multirow{2}{*}{PCA}   & 0.656 & 0.320 & 0.522 & \textbf{0.651} & 0.348 & 0.530 & \textbf{0.656} & 0.320 & 0.522 & \textbf{0.651} & 0.348 & \textbf{0.530} \\
& \pmS{0.009} & \pmS{0.005} & \pmS{0.005} & \pmS{0.019} & \pmS{0.000} & \pmS{0.012} & \pmS{0.000} & \pmS{0.002} & \pmS{0.001} & \pmS{0.000} & \pmS{0.001} & \pmS{0.000} \\
\bottomrule
\end{tabular}
\label{tab:unimodal_dsbn}

\end{table*}

\begin{table}[ht!]
\vspace{-0.5cm}
\centering
\caption{Comparison of different multi-omics integration methods using the CLEAR pipeline. Data integration metrics were computed for the BMMC dataset.}
\vspace{0.3cm}
\setlength\tabcolsep{10pt}
\begin{tabular}{l|ccc|ccc|ccc}
\toprule
\multirow{2}{*}{\textbf{Method}} & \multicolumn{3}{c|}{\textbf{Add}} & \multicolumn{3}{c|}{\textbf{Concat}} & \multicolumn{3}{c}{\textbf{CLIP + Concat}}  \\
& Bio & Batch & Total & Bio & Batch & Total & Bio & Batch & Total \\
\midrule

\multirow{2}{*}{SimCLR} & 0.827       & 0.3         & 0.616    & 0.84       & 0.273      & 0.613                 & 0.511      & \textbf{0.504}      &  0.508           \\
                        & \pmS{0.078} & \pmS{0.057} & \pmS{0.05} & \pmS{0.093}& \pmS{0.058}& \pmS{0.065}           & \pmS{0.223}& \pmS{0.094}&  \pmS{0.166}     \\
\multirow{2}{*}{MoCo}   & \textbf{0.935}       & 0.407       & \textbf{0.724}     & 0.056      & \textbf{0.8}        & 0.354                 & 0.566      & 0.464      &   0.525          \\
                        & \pmS{0.07}  & \pmS{0.019} & \pmS{0.045}& \pmS{0.065}& \pmS{0.000}& \pmS{0.039}           & \pmS{0.157}& \pmS{0.161}&  \pmS{0.049}     \\
\multirow{2}{*}{SimSiam}& 0.453       & 0.174       & 0.341               & 0.506      & 0.21       & 0.387       & 0.197      & 0.364      &  0.264           \\
                        & \pmS{0.175} & \pmS{0.025} & \pmS{0.107}         & \pmS{0.146}& \pmS{0.041}& \pmS{0.083}  & \pmS{0.177}& \pmS{0.051}&  \pmS{0.011}     \\
\multirow{2}{*}{NNCLR}  & 0.679       & 0.225       & 0.498               & 0.768      & 0.231      & 0.553       & 0.584      & 0.5        &  0.551           \\
                        & \pmS{0.171} & \pmS{0.041} & \pmS{0.113}         & \pmS{0.105}& \pmS{0.034}& \pmS{0.06}  & \pmS{0.088}& \pmS{0.079}&  \pmS{0.066}     \\
\multirow{2}{*}{BYOL}   & 0.117       & \textbf{0.8}         & 0.39                & 0.527      & 0.673      & 0.586       & 0.46       & 0.403      &  0.437           \\
                        & \pmS{0.109} & \pmS{0.000} & \pmS{0.066}         & \pmS{0.029}& \pmS{0.074}& \pmS{0.03}  & \pmS{0.123} & \pmS{0.152}&  \pmS{0.115}     \\
\multirow{2}{*}{VICReg} & 0.791       & 0.449       & 0.654               & \textbf{0.887}      & 0.484      & \textbf{0.726}      & \textbf{0.72}       & 0.38       &  \textbf{0.584}           \\
                        & \pmS{0.089} & \pmS{0.014} & \pmS{0.052}         & \pmS{0.035}& \pmS{0.009}& \pmS{0.022}  & \pmS{0.079}& \pmS{0.055}&  \pmS{0.056}     \\
Barlow                  & 0.717       & 0.262       & 0.535               & 0.852      & 0.27       & 0.62        & 0.706      & 0.352      &  0.565           \\
Twins                   & \pmS{0.055} & \pmS{0.026} & \pmS{0.039}         & \pmS{0.096}& \pmS{0.012}& \pmS{0.059}  & \pmS{0.115}& \pmS{0.085}&  \pmS{0.055}     \\

\bottomrule
\end{tabular}

\label{tab:Neurips_compare_add_concat_clip_bc}

\vspace{-1cm}

\end{table}

\begin{table}[ht!]
\setlength\tabcolsep{17.8pt}
\centering
\caption{Augmentation Parameters for the CLEAR~\cite{CLEAR} augmentations on the left. Results for the ablation of all augmentations on the right, including the CLAIRE~\cite{XuhuaRFL23} augmentation denoted as MNN, and our BBKNN augmentation. Results stem from our ablation detailed in~\autoref{sec:eval_details}.}
\vspace{0.3cm}
\begin{tabular}{l|ccc|ccc}
\toprule
\multirow{2}{*}{\textbf{Augmentation}} & \multicolumn{3}{|c|}{\textbf{CLEAR}} & \multicolumn{3}{|c}{\textbf{Ablation Result}}\\
& $\alpha$ & $\sigma$ & \textit{knn} & $\alpha$ & $\sigma$ & \textit{knn}\\
\midrule
Masking & 0.2 & --- & --- & 0.5 & --- & ---\\
Gaussian Noise & 0.8 & 0.2 & --- & 0.3 & 0.2 & ---\\
InnerSwap & 0.1 & --- & --- & 0.3 & --- & --- \\
CrossOver & 0.25 & --- & --- & 0.1 & --- & ---\\
BBKNN & --- & --- & --- & 0.9 & --- & 3 \\
MNN & --- & --- & --- & 0.5 & --- & 3 \\
\bottomrule
\end{tabular}
\label{tab:augmentation-parameters}
\vspace{-0.5cm}
\end{table}

\begin{table}[ht!]
\centering
\caption{Batch correction benchmark for the Tabula Sapiens dataset (1.1 million cells) trained using the CLEAR pipeline (only 1 run) to show the scalability of our benchmark. The previously drawn conclusion that baselines outperform SSL methods in uni-modal batch correction holds. Generic SSL methods are good at batch correction but not preservation of the true biological variance.}
\vspace{0.3cm}
\setlength\tabcolsep{30pt}
\begin{tabular}{l|ccc}
\toprule
\multirow{2}{*}{\textbf{Method}} & \multicolumn{3}{c}{\textbf{Tabula Sapiens}} \\
& Bio & Batch & Total \\
\midrule
SimCLR  & 0.374 & 0.567 & 0.451 \\
MoCo    & 0.342 & 0.541 & 0.421 \\
SimSiam & 0.237 & 0.430 & 0.314 \\
NNCLR   & 0.312 & 0.454 & 0.368 \\
BYOL    & 0.085 & \textbf{0.800} & 0.371 \\
VICReg  & 0.327 & 0.657 & 0.459 \\
BarlowTwins & 0.376 & 0.427 & 0.396 \\
\midrule
scVI    & \textbf{0.723} & 0.254 & \textbf{0.536} \\
\midrule
PCA     & 0.538 & 0.297 & 0.442 \\
\bottomrule
\end{tabular}
\label{tab:big_sapiens_bc}
\end{table}


\end{document}